\documentclass[sigconf,screen,nonacm]{acmart}

\setcopyright{none}
\usepackage{amsmath}
\usepackage{amsfonts}
\usepackage{algorithmic}
\usepackage{graphicx}
\usepackage{textcomp}
\usepackage{xcolor}
\usepackage{cleveref}
\usepackage{enumitem}
\usepackage{booktabs}
\usepackage{multirow}
\usepackage{makecell}

\newcommand{\figOne}{
\begin{figure}[t]
    \centering
    \includegraphics[width=\linewidth]{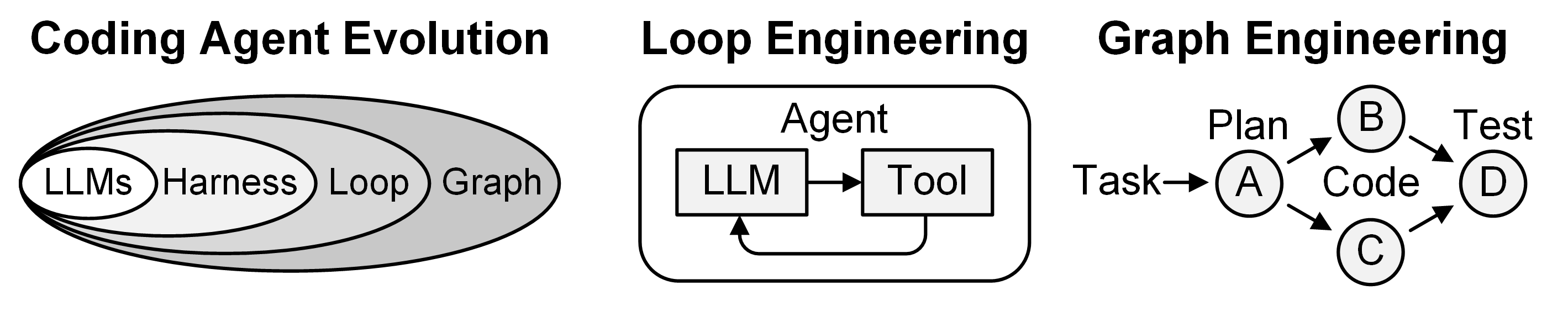}
    \caption{Coding Agent workflows: From LLMs to Loop/Graph Engineering.}
    \label{fig:figOne}
    \vspace{-0.5cm}
\end{figure}
}

\newcommand{\figTwo}{
\begin{figure}[t]
    \centering
    \includegraphics[width=\linewidth]{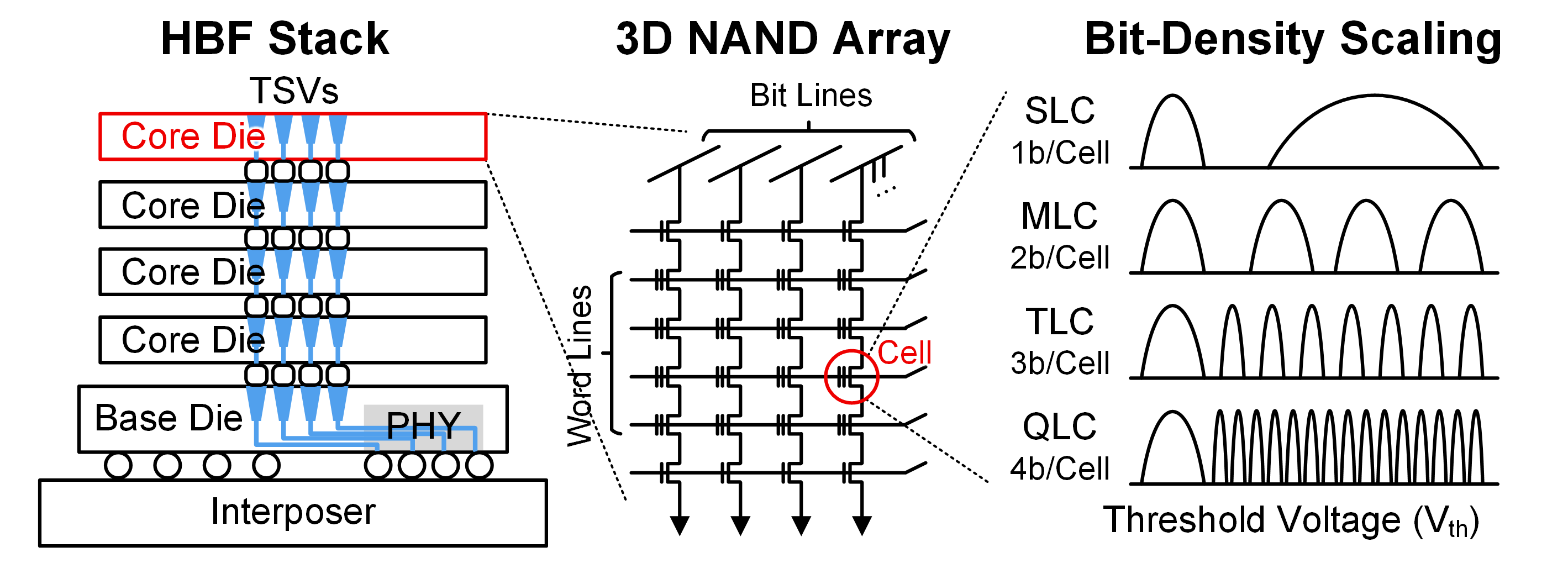}
    \caption{Physical organization of HBF stack/3D NAND array, and bit-density scaling from SLC to QLC.}
    \label{fig:figTwo}
    \vspace{-0.5cm}
\end{figure}
}

\newcommand{\figThree}{
\begin{figure}[t]
    \centering
    \includegraphics[width=\linewidth]{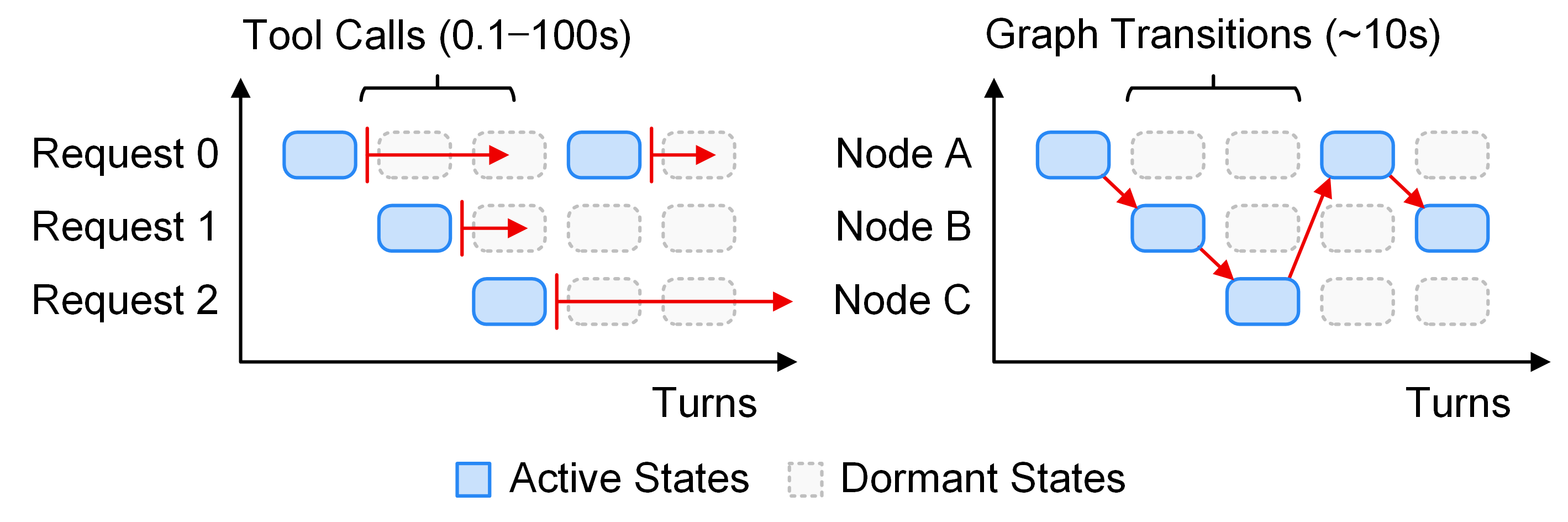}
    \caption{State amplification caused by active/dormant states in Loop Engineering (left) and Graph Engineering (right).}
    \label{fig:figThree}
    \vspace{-0.5cm}
\end{figure}
}

\newcommand{\figFour}{
\begin{figure}[t]
    \centering
    \includegraphics[width=\linewidth]{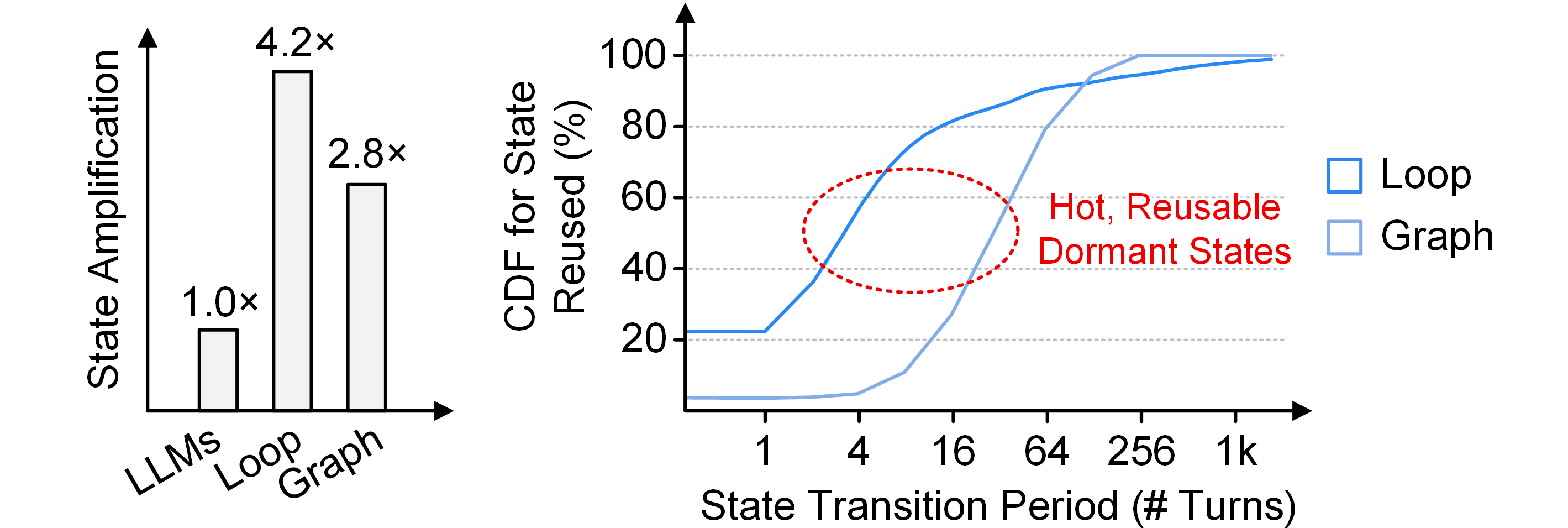}
    \caption{State amplification in Coding Agent (left) and cumulative state transition period distributions (right) in GLM-5.1.}
    \label{fig:figFour}
    \vspace{-0.5cm}
\end{figure}
}

\newcommand{\figFive}{
\begin{figure}[t]
    \centering
    \includegraphics[width=\linewidth]{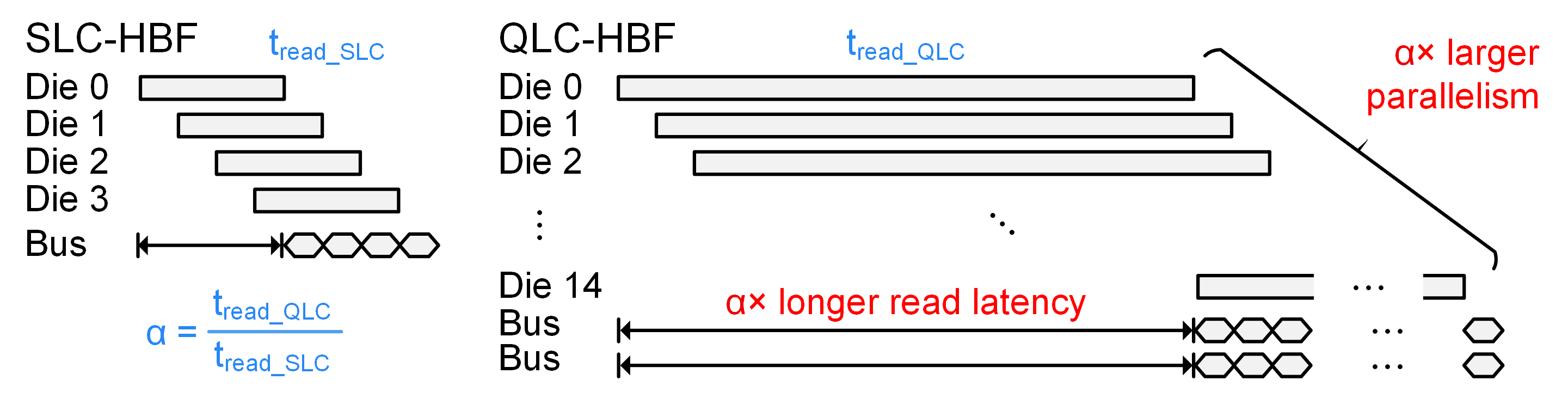}
    \caption{Pipeline diagram of interleaved NAND reads for latency hiding in SLC- and QLC-HBF.}
    \label{fig:figFive}
    \vspace{-0.5cm}
\end{figure}
}

\newcommand{\figSix}{
\begin{figure}[t]
    \centering
    \includegraphics[width=\linewidth]{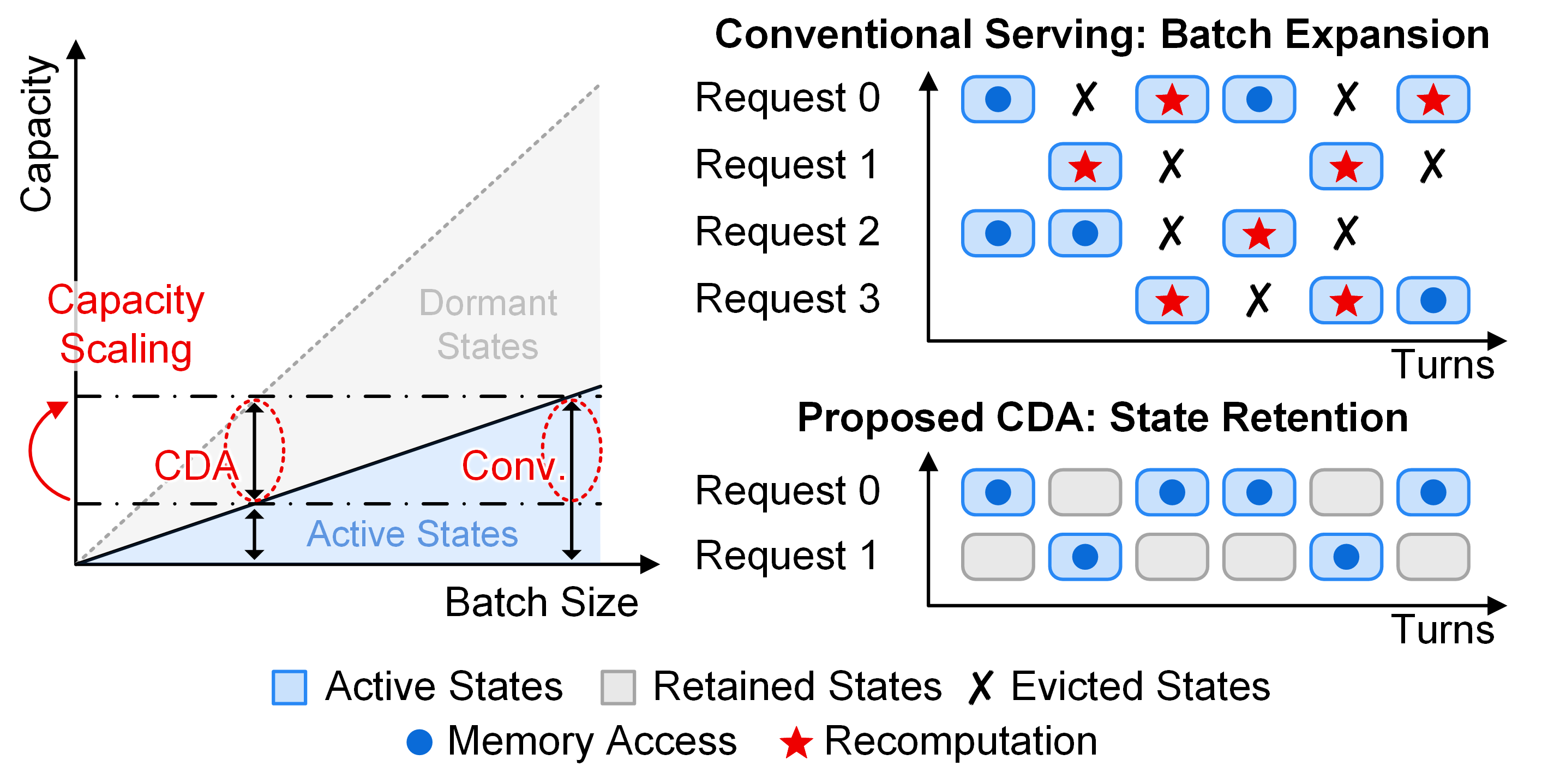}
    \caption{Comparison of memory capacity allocation strategies: batch expansion with dormant-state eviction and recomputation versus CDA with KV cache retention and retrieval.}
    \label{fig:figSix}
    \vspace{-0.5cm}
\end{figure}
}

\newcommand{\figSeven}{
\begin{figure}[t]
    \centering
    \includegraphics[width=\linewidth]{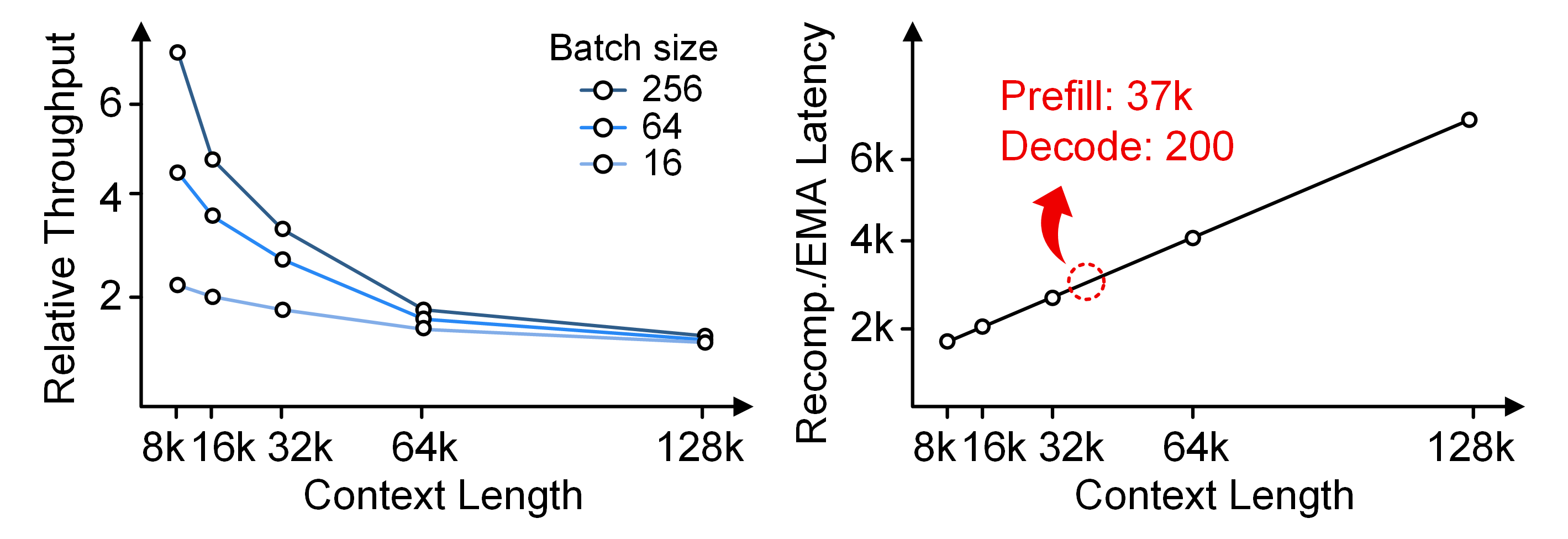}
    \caption{Diminishing throughput benefits of BE with longer contexts (left) and increasing recomputation-to-EMA latency (right) for Qwen3-Coder-30B-A3B on SWE-bench Verified.}
    \label{fig:figSeven}
    \vspace{-0.5cm}
\end{figure}
}

\newcommand{\figEight}{
\begin{figure}[t]
    \centering
    \includegraphics[width=\linewidth]{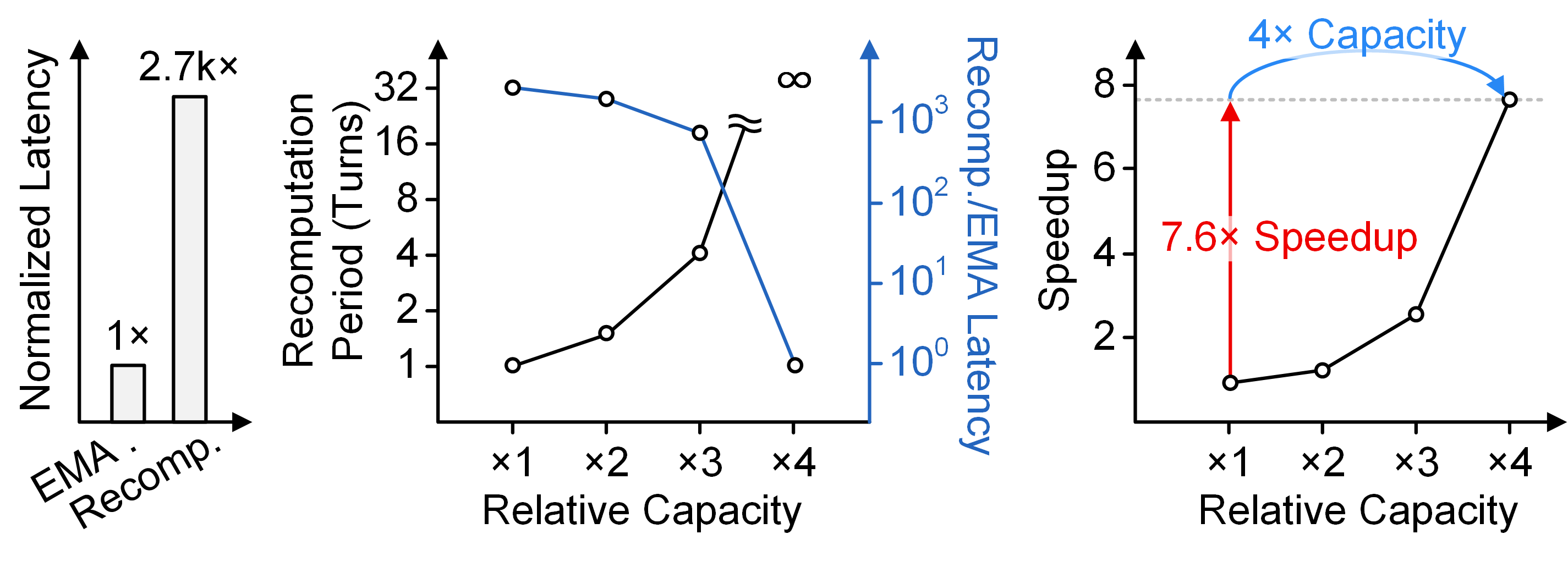}
    \caption{Recomputation versus EMA hardware cost (left), extension of recomputation intervals with increased retention capacity (middle), and CDA cost reduction through capacity scaling (right) in Qwen3-Coder-30B-A3B.}
    \label{fig:figEight}
    \vspace{-0.5cm}
\end{figure}
}

\newcommand{\figNine}{
\begin{figure}[t]
    \centering
    \includegraphics[width=\linewidth]{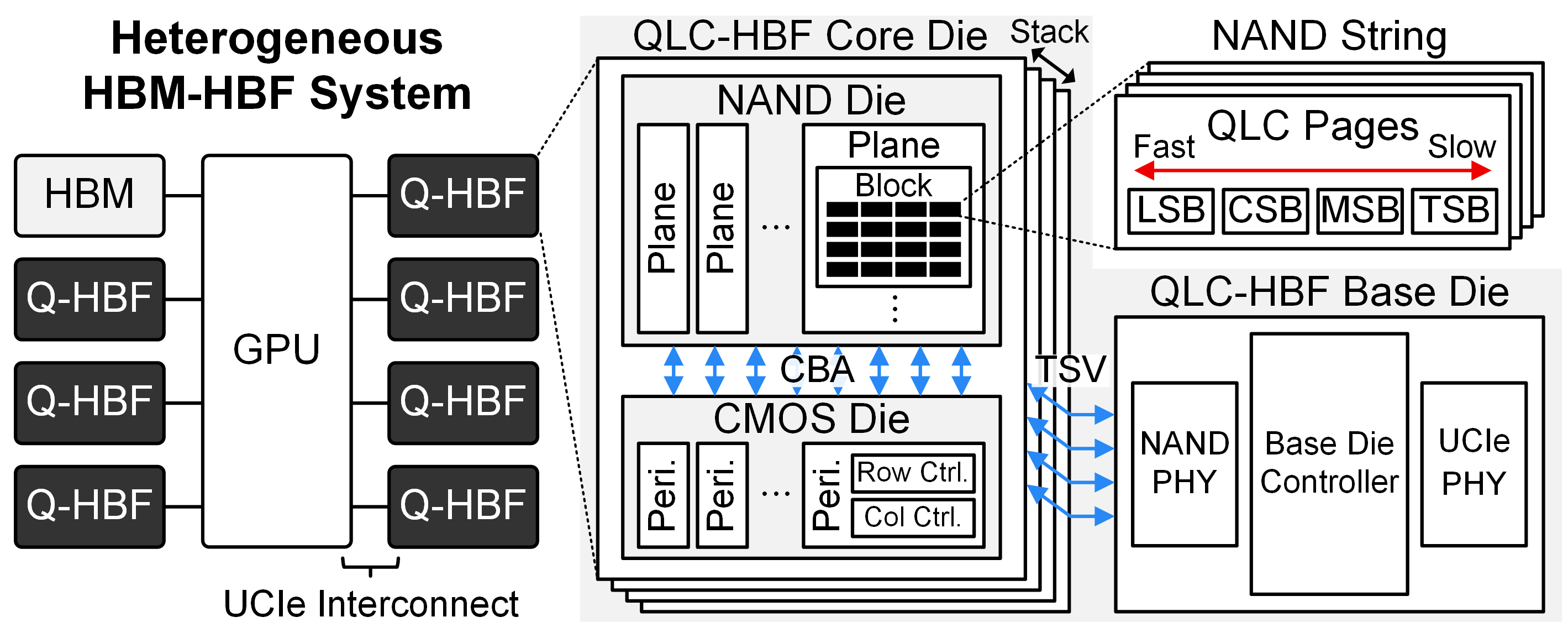}
    \caption{Overall LaFlash architecture.}
    \label{fig:figNine}
    \vspace{-0.5cm}
\end{figure}
}

\newcommand{\figTen}{
\begin{figure}[t]
    \centering
    \includegraphics[width=\linewidth]{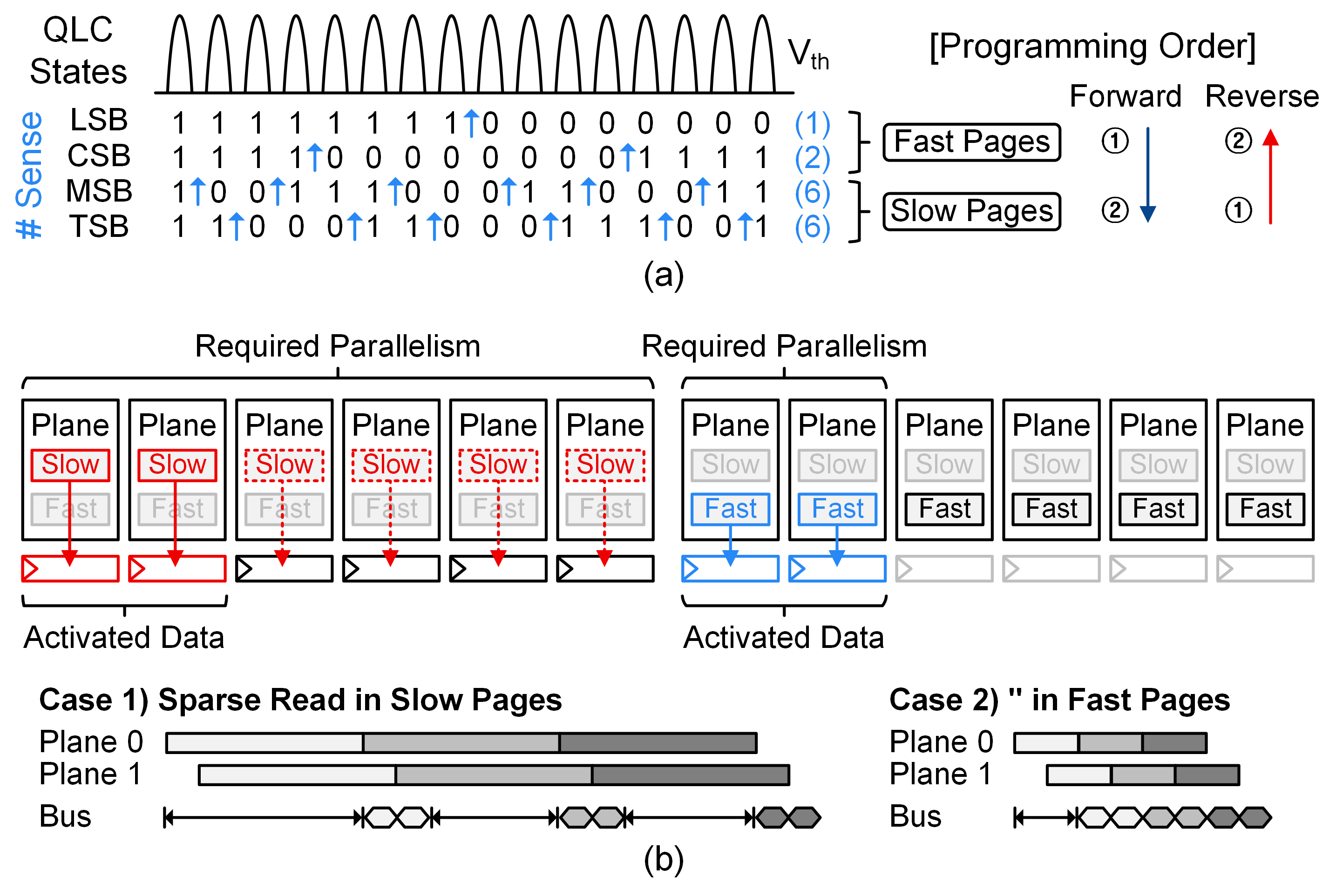}
    \caption{QLC page-dependent sensing and programming (a), and fast-page allocation to improve bandwidth utilization under sparse accesses (b).}
    \label{fig:figTen}
    \vspace{-0.5cm}
\end{figure}
}

\newcommand{\figEleven}{
\begin{figure}[t]
    \centering
    \includegraphics[width=\linewidth]{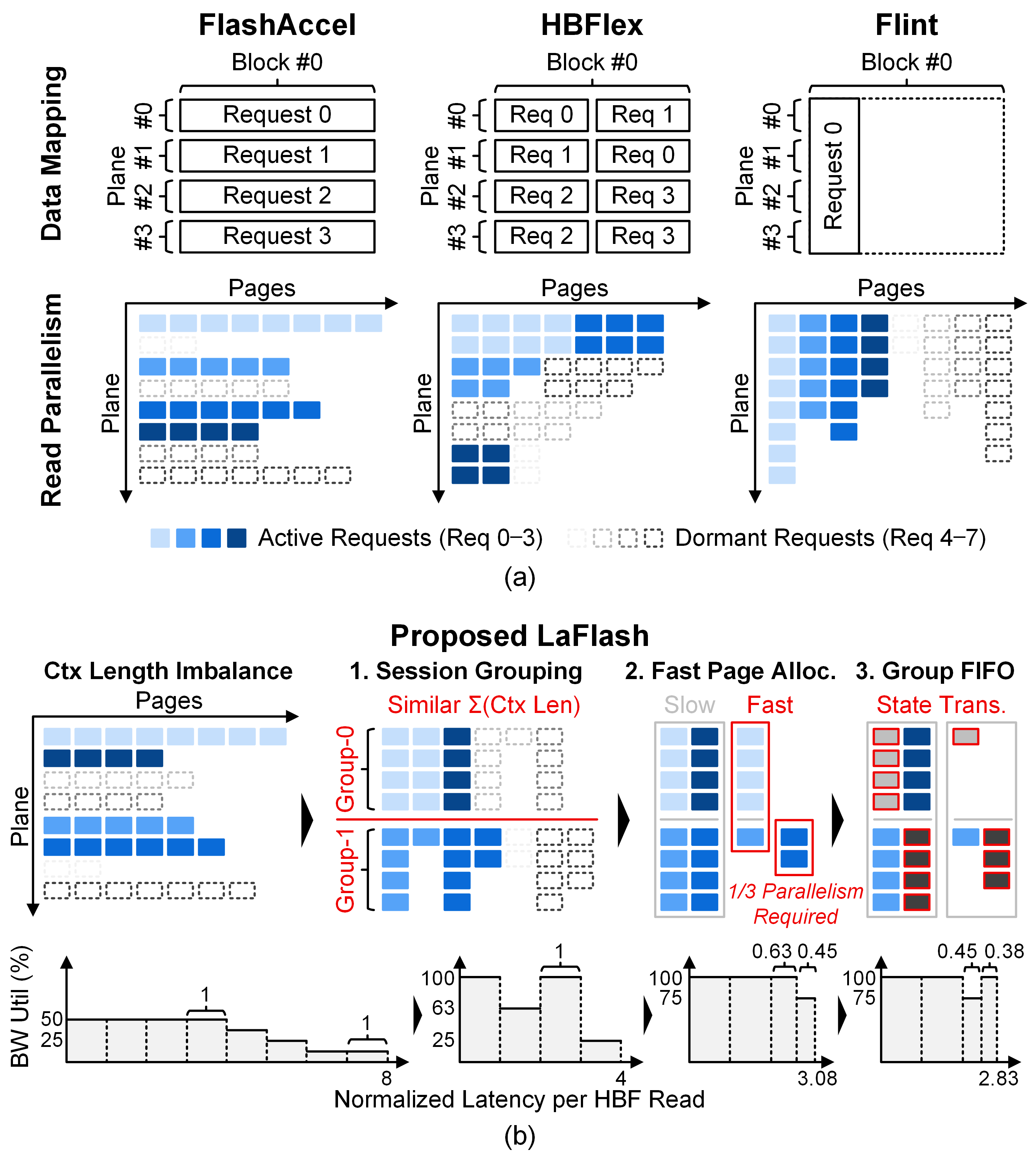}
    \caption{Limitations of conventional HBF mapping (a) and LaFlash's Request Grouping and Fast-Page Allocation (b).}
    \label{fig:figEleven}
    \vspace{-0.5cm}
\end{figure}
}

\newcommand{\figTwelve}{
\begin{figure}[t]
    \centering
    \includegraphics[width=\linewidth]{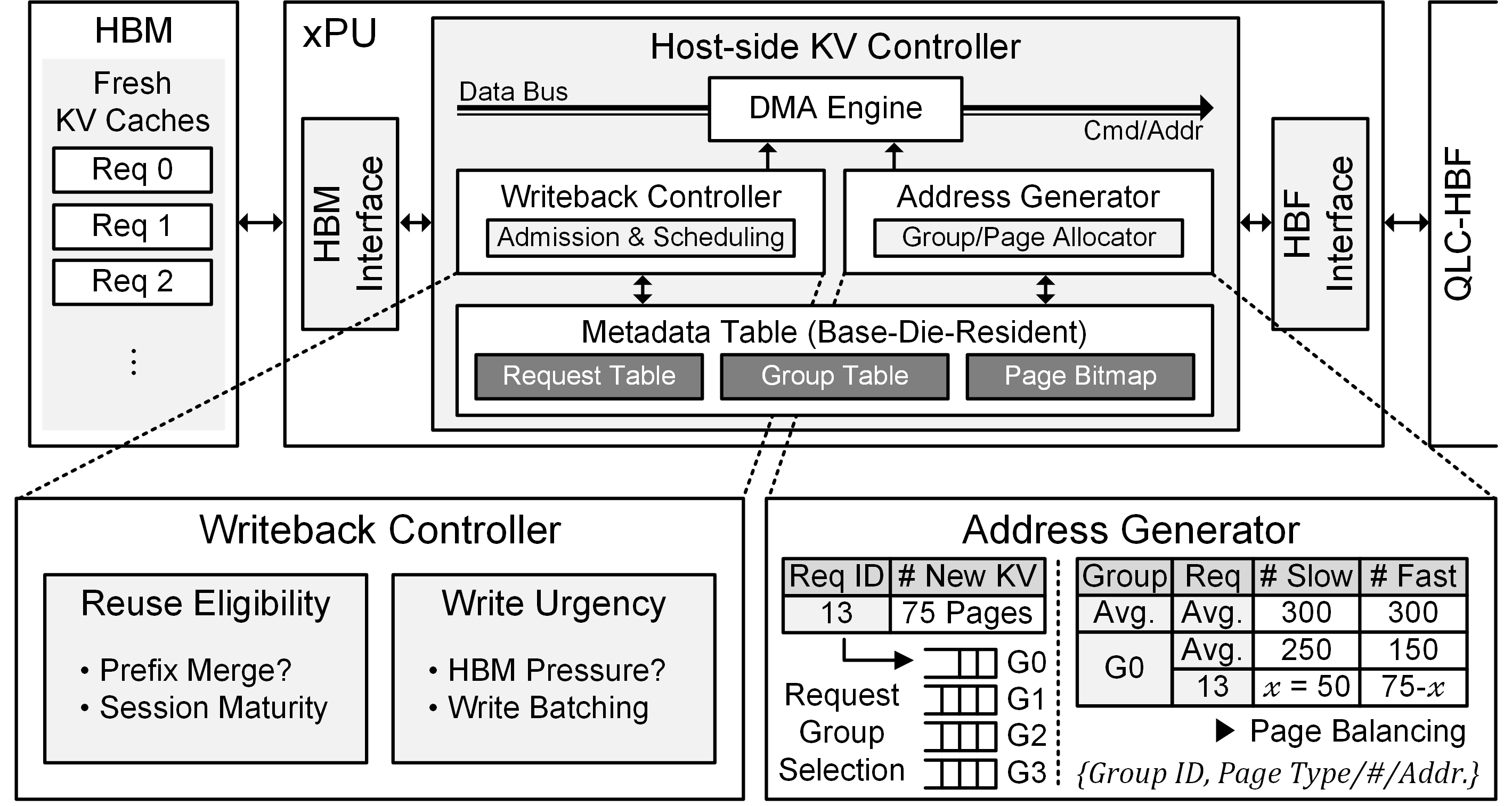}
    \caption{LaFlash host-side KV controller architecture.}
    \label{fig:figTwelve}
    \vspace{-0.5cm}
\end{figure}
}

\newcommand{\figThirteen}{
\begin{figure}[t]
    \centering
    \includegraphics[width=\linewidth]{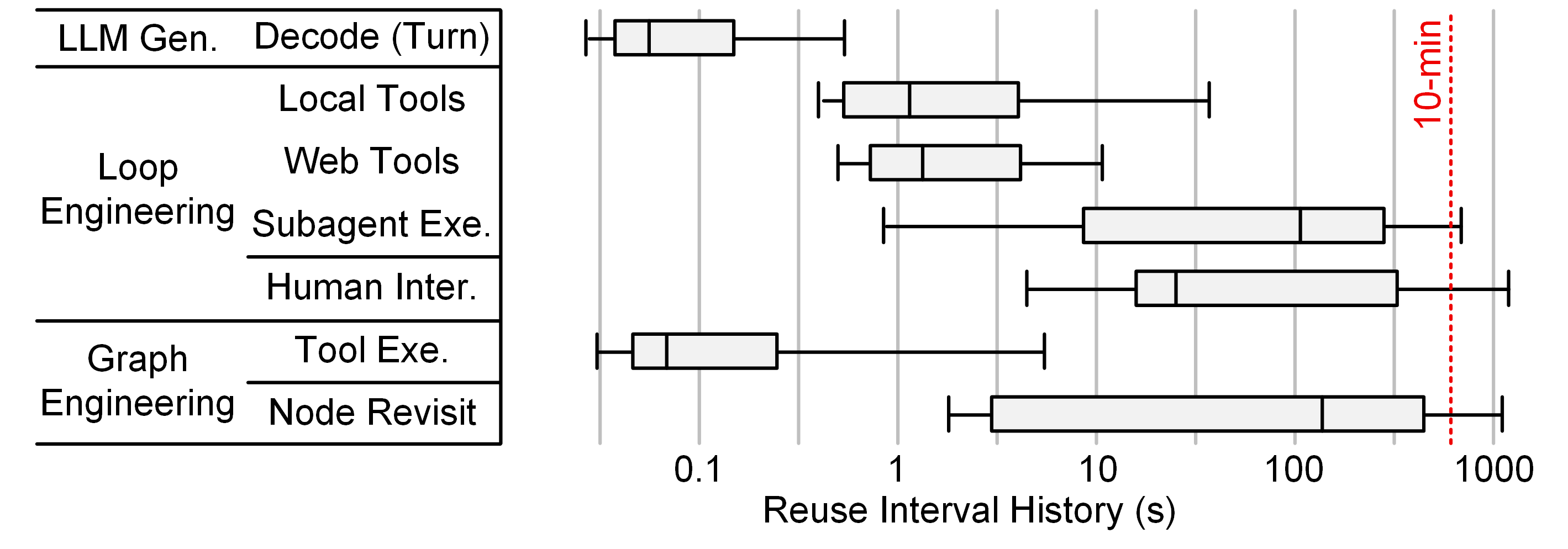}
    \caption{State dormancy and reuse interval distributions in Coding Agent workloads using MiniMax-M2.7 (P5--P95 whiskers, P25--P75 boxes, and P50 median).}
    \label{fig:figThirteen}
    \vspace{-0.5cm}
\end{figure}
}

\newcommand{\figFourteen}{
\begin{figure}[t]
    \centering
    \includegraphics[width=\linewidth]{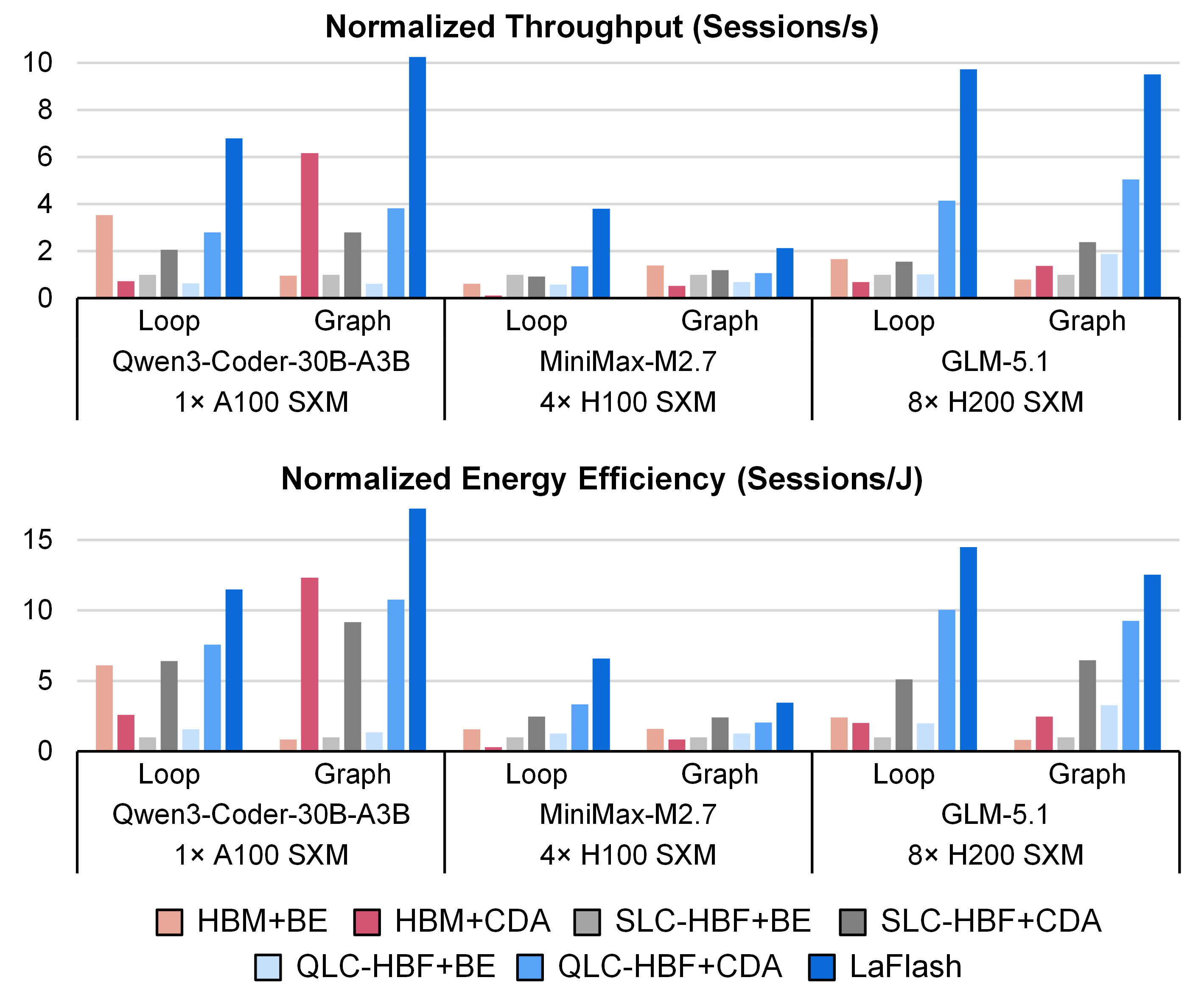}
    \caption{End-to-end throughput (top) and energy efficiency (bottom) of LaFlash across Coding Agent workloads and GPU–LLM configurations with active concurrency $N$.}
    \label{fig:figFourteen}
    \vspace{-0.5cm}
\end{figure}
}

\newcommand{\figFifteen}{
\begin{figure}[t]
    \centering
    \includegraphics[width=\linewidth]{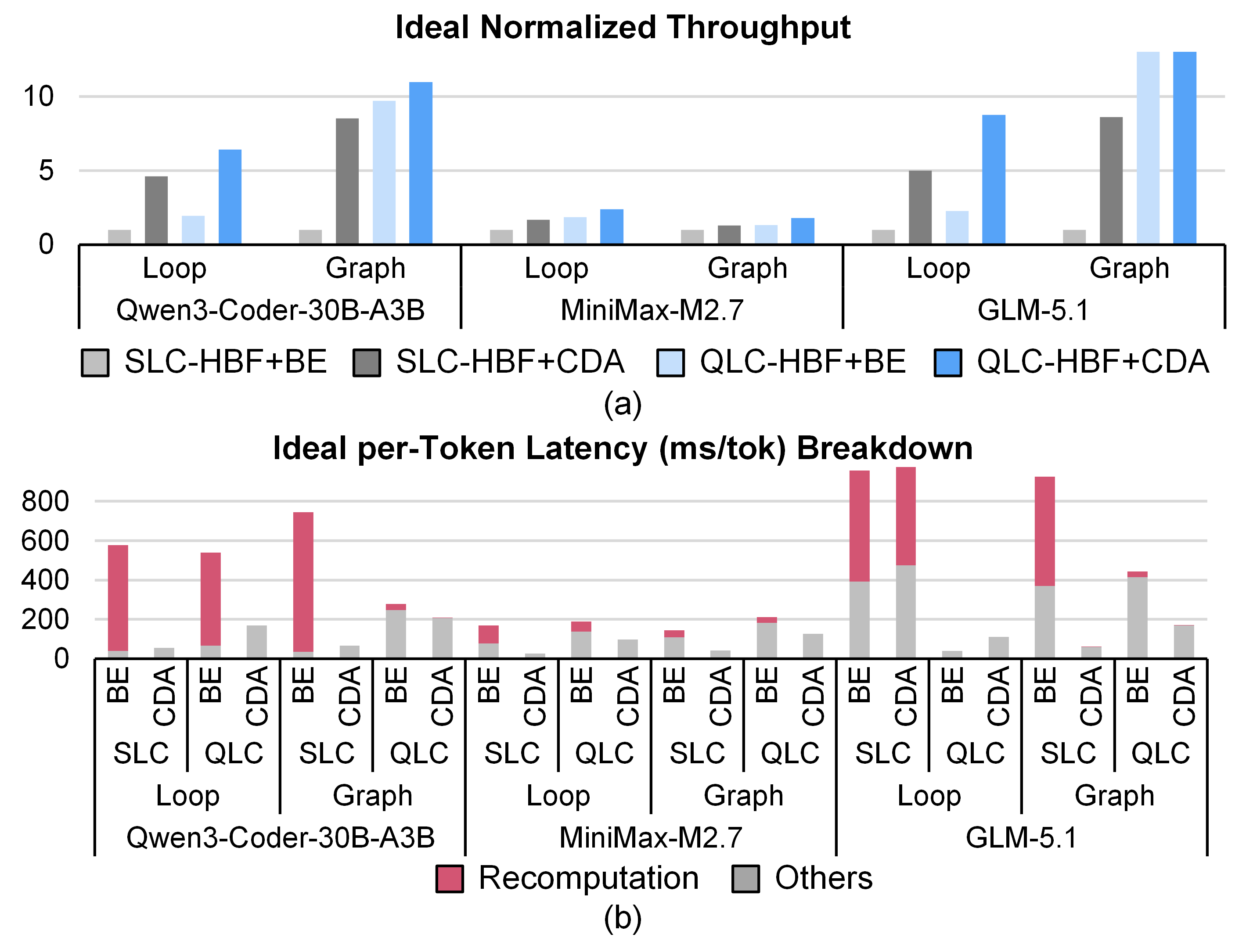}
    \caption{Throughput gains from capacity expansion without hardware constraints (a) and latency breakdown including recomputation overhead (b).}
    \label{fig:figFifteen}
    \vspace{-0.5cm}
\end{figure}
}

\newcommand{\figSixteen}{
\begin{figure}[t]
    \centering
    \includegraphics[width=\linewidth]{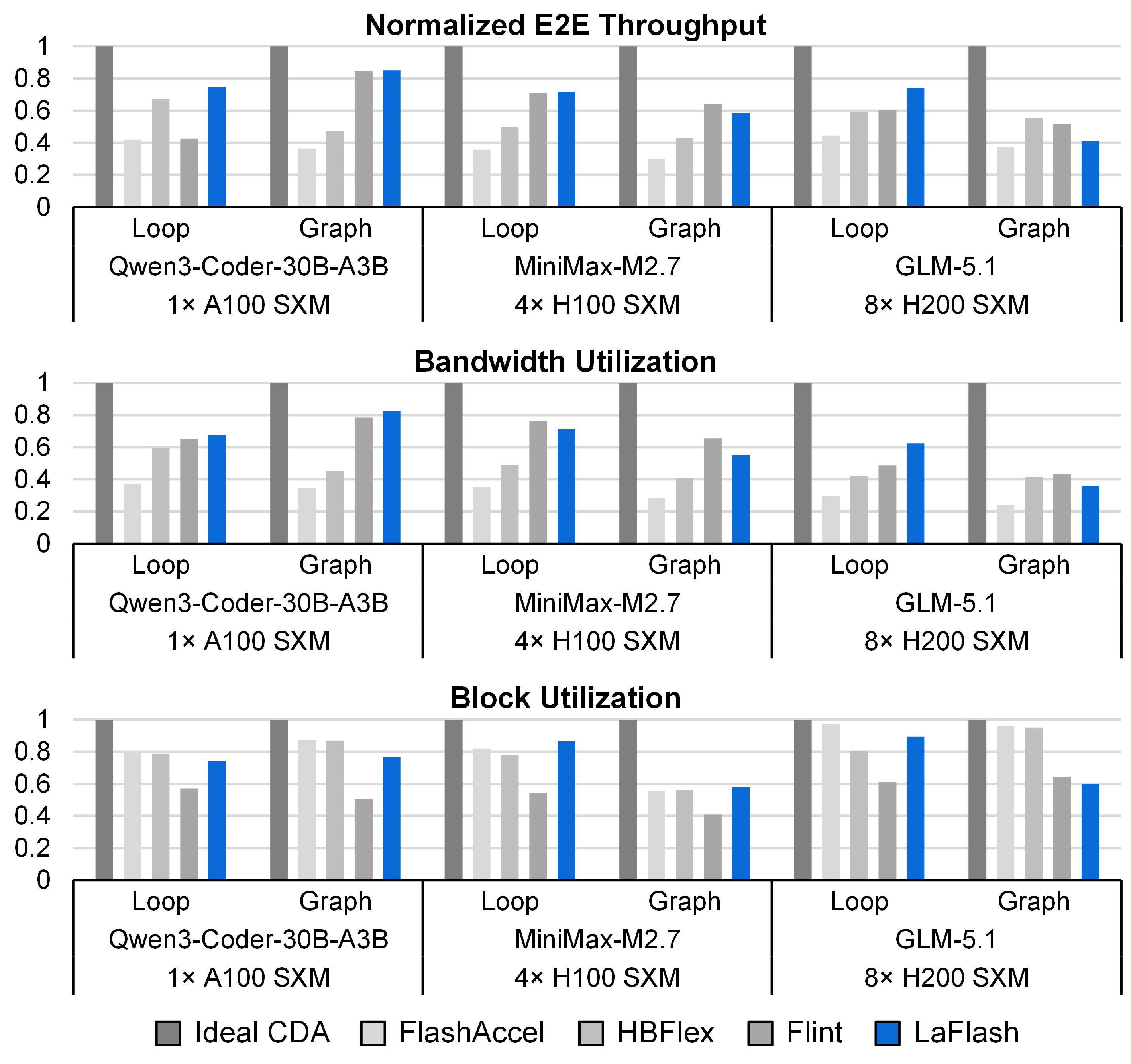}
    \caption{Comparison of QLC-HBF mapping schemes with a session group size of 128 planes.}
    \label{fig:figSixteen}
    \vspace{-0.5cm}
\end{figure}
}

\newcommand{\figSeventeen}{
\begin{figure}[t]
    \centering
    \includegraphics[width=\linewidth]{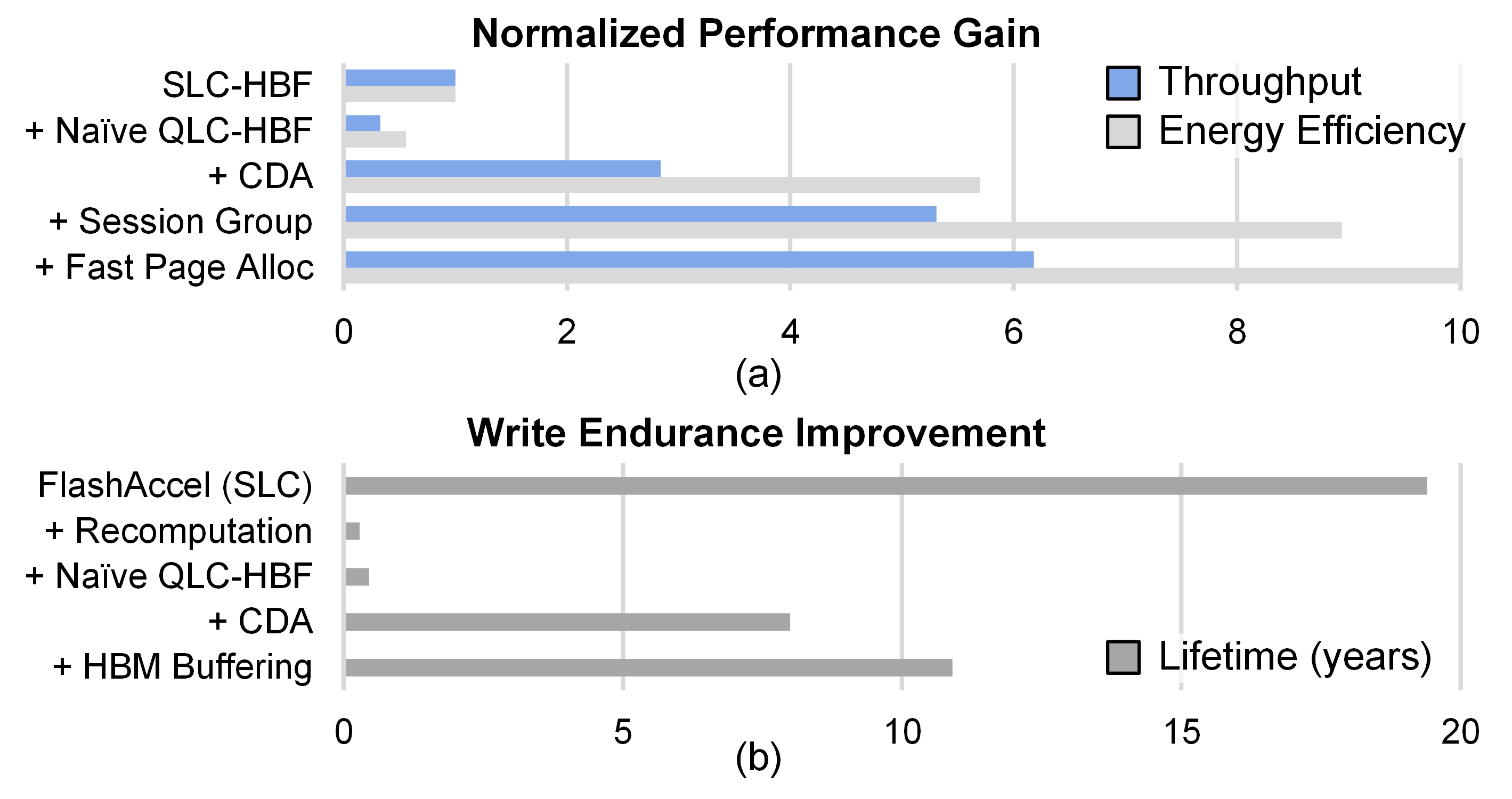}
    \caption{Ablation of LaFlash's performance (a) and write-endurance optimizations (b) using MiniMax-M2.7 on Loop Engineering workloads from FreeInference and Pi-Trace.}
    \label{fig:figSeventeen}
    \vspace{-0.5cm}
\end{figure}
}

\newcommand{\figEighteen}{
\begin{figure}[t]
    \centering
    \includegraphics[width=\linewidth]{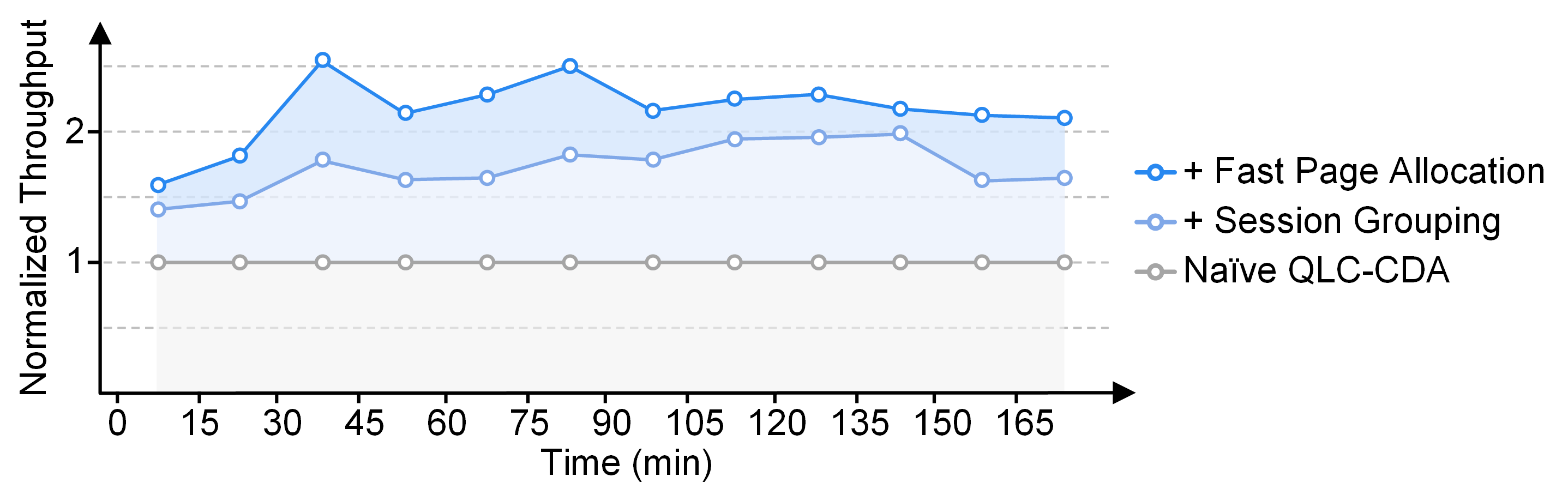}
    \caption{Cumulative throughput gains of LaFlash over Naive QLC under realistic vLLM continuous batching for MiniMax-M2.7 Loop Engineering.}
    \label{fig:figEighteen}
    \vspace{-0.5cm}
\end{figure}
}

\newcommand{\tabOne}{
\begin{table}[t]
\centering
\footnotesize
\setlength{\tabcolsep}{1.8pt}
\renewcommand{\arraystretch}{1.08}
\caption{Hardware specifications, workloads, and
deployment configurations of the evaluated GPU--HBF platforms.}
\label{tab:tabOne}
\begin{tabular*}{\linewidth}{@{\extracolsep{\fill}}llccc}
\toprule
\textbf{Config.} & \textbf{Parameter} &
\textbf{A100} & \textbf{H100} & \textbf{H200} \\
\midrule
\multirow{2}{*}{GPU}
& Peak BF16 (TFLOPS/GPU) & 312 & 989 & 989 \\
& Peak FP8 (TFLOPS/GPU) & -- & 1,979 & 1,979 \\
\midrule
\multirow{3}{*}{HBM}
& Stacks/GPU & 5 & 5 & 6 \\
& Capacity/Stack (GB) & 16 & 16 & 23.5 \\
& BW/Stack (GB/s) & 408 & 670 & 800 \\
\midrule
\multirow{3}{*}{HBF}
& SLC Capacity/Stack (GB) & 128 & 128 & 188 \\
& QLC Capacity/Stack (GB) & 512 & 512 & 752 \\
& BW/Stack (GB/s) & 512 & 768 & 1024 \\
\midrule
\multirow{5}{*}{System}
& HBM:HBF Stacks/GPU & 1:4 & 1:4 & 2:4 \\
& HBM-only Capacity (GB) & 80 & 80 & 141 \\
& SLC-Hybrid Capacity (GB) & 528 & 528 & 799 \\
& QLC-Hybrid Capacity (GB) & 2,064 & 2,064 & 3,055 \\
& Peak BW/GPU (TB/s) & 2.46 & 3.74 & 5.70 \\
\midrule
\multirow{6}{*}{Model}
& Model & Qwen3-Coder & MiniMax-M2.7 & GLM-5.1 \\
& Total Params (B) & 30.5 & 229 & 754 \\
& Active Params (B) & 3.3 & 10 & 40 \\
& Attention Type & GQA & GQA & MLA + DSA \\
& Transformer Layers & 48 & 62 & 78 \\
& Data Precision & FP8 & FP8 & FP8 \\
\midrule
\multirow{3}{*}{Deploy.}
& GPU Count & 1 & 4 & 8 \\
& Attention Parallelism & None & TP=4 & DP=8 \\
& MoE FFN Parallelism & None & EP=4 & EP=8 \\
\bottomrule
\end{tabular*}
\vspace{-2mm}
\end{table}
}

\newcommand{\tabTwo}{
\begin{table}[t]
\centering
\footnotesize
\setlength{\tabcolsep}{3pt}
\renewcommand{\arraystretch}{1.12}
\caption{System specifications
of the evaluated SLC- and QLC-HBF architectures on NVIDIA H100 SXM.}
\label{tab:tabTwo}
\begin{tabular*}{\linewidth}{@{\extracolsep{\fill}}llcc}
\toprule
\textbf{Category} & \textbf{Parameter} &
\textbf{SLC-HBF} & \textbf{QLC-HBF} \\
\midrule

\multirow{5}{*}{\textbf{Config.}}
& Bits/Cell & 1 & 4 \\
& Capacity/Stack (GB) & 128 & 512 \\
& Bandwidth/Stack (GB/s) & 768 & 768 \\
& Total Planes/Stack & 768 & 2,688 \\
& P/E Endurance (cycles) & 100K & 1K \\

\midrule
\multirow{5}{*}{\textbf{Organization}}
& Page Size (KiB) & 4 & 4 \\
& Pages/Block & 256 & 1,024 \\
& Blocks/Plane & 159 & 45 \\
& Planes/Die & 96 & 336 \\
& Dies/Stack & 8 & 8 \\

\midrule
\multirow{4}{*}{\textbf{Timing}}
& Sensing Counts/Page & 1 & 1/2/6/6 \\
& Read Latency ($\mu$s) & 4 & 4/6/14/14 \\
& Program Latency ($\mu$s) & 75 & \makecell{250/6,270 (RTSP) \\ 1,000/5,520 (FTSP)} \\
& Erase Latency (ms) & 2 & 10 \\

\midrule
\multirow{2}{*}{\textbf{Energy}}
& Read Energy (pJ/bit) & 1.01 & 1.01 / 1.50 / 3.51 / 3.51 \\
& Write Energy (pJ/bit) & 18.62 & 101.47 \\

\bottomrule
\end{tabular*}
\vspace{-2mm}
\end{table}
}

\newcommand{\tabThree}{
\begin{table}[t]
\centering
\footnotesize
\setlength{\tabcolsep}{1.8pt}
\renewcommand{\arraystretch}{1.08}
\caption{Workflow configurations and workload characteristics
of the evaluated Coding Agent workloads.}
\label{tab:tabThree}
\begin{tabular*}{\linewidth}{@{\extracolsep{\fill}}lcc}
\toprule
\textbf{Parameter} &
\textbf{Loop Engineering} & \textbf{Graph Engineering} \\
\midrule
\multicolumn{3}{c}{\textbf{Workflow Configuration}} \\
\midrule
Benchmark & FreeInference, $\pi$ & Terminal-Bench 2.1 / SWE-bench Pro  \\
Agent Framework & Claude, Hermes, etc. & Claude/Hermes+LangGraph \\
LLM Backbone &
\multicolumn{2}{c}{Qwen3-Coder-30B-A3B, MiniMax-M2.7, GLM-5.1} \\
Serving Engine & \multicolumn{2}{c}{vLLM} \\
\midrule
\multicolumn{3}{c}{\textbf{Generation Characteristics (Avg./P50/P90/P99)}} \\
\midrule
Context Length/Session & 96k/91k/174k/199k & 116k/107k/185k/264k \\
Context Length/Node & -- & 48k/43k/74k/127k \\
Prefill Tokens/Turn & 92k/93k/151k/187k & 42k/38k/74k/122k \\
Decode Tokens/Turn & 398/164/906/3.7k & 373/168/852/2.9k \\
Tool Results/Turn & 595/83/1.2k/10k & 302/40/625/4.4k \\
Turns/Session & 80/34/193/628 & 148/140/251/341 \\
\midrule
\multicolumn{3}{c}{\textbf{Execution Characteristics (Avg./P50/P90/P99)}} \\
\midrule
Tool Calls/Turn & 1.05/1/1/4 & 1.24/1/2/6 \\
Tool Latency (s) & 20/1.3/12.1/168 & 1.7/0.04/0.7/37.5 \\
Graph Nodes/Task & -- & 4/--/--/-- \\
Turns/Node & -- & 37/31/70/122 \\
\bottomrule
\end{tabular*}
\vspace{-2mm}
\end{table}
}

\newcommand{\tabFour}{
\begin{table}[t]
\centering
\footnotesize
\setlength{\tabcolsep}{2pt}
\renewcommand{\arraystretch}{1.12}
\caption{Recomputation versus EMA unit cost (latency) comparison across context lengths and GPU--LLM configurations.}
\label{tab:tabFour}
\begin{tabular*}{\linewidth}{@{\extracolsep{\fill}}llcccc}
\toprule
\multirow{2}{*}{\textbf{GPU}} &
\multirow{2}{*}{\textbf{Model}} &
\multicolumn{4}{c}{\textbf{Context Length (Tokens)}} \\
\cmidrule(lr){3-6}
& & \textbf{50K} & \textbf{100K} &
\textbf{150K} & \textbf{200K} \\
\midrule
1$\times$ A100 SXM & Qwen3-Coder-30B-A3B & 3,359 & 5,981 & 8,606 & 11,232 \\
4$\times$ H100 SXM & MiniMax-M2.7 & 1,434 & 1,881 & 2,340 & 2,803 \\
8$\times$ H200 SXM & GLM-5.1 & 4,362 & 4,972 & 5,580 & 6,188 \\
\bottomrule
\end{tabular*}
\vspace{-2mm}
\end{table}
}

\newcommand{\tabConcurrency}{
\begin{table*}[t]
\centering
\footnotesize
\setlength{\tabcolsep}{2pt}
\renewcommand{\arraystretch}{1.12}
\caption{Concurrency determination with state amplification (SA)
and memory pressure ($\rho=0.5$). $N_M$ denotes memory-limited
concurrency, with the binding constraint in bold.
GLM-5.1 is reported per DP rank.}
\label{tab:concurrency}
\resizebox{\textwidth}{!}{%
\begin{tabular}{llcccccccccccc}
\toprule
\multicolumn{2}{c}{\textbf{Workflow}} &
\multirow{2}{*}{\shortstack{\textbf{KV/Token}\\\textbf{(KiB)}}} &
\multirow{2}{*}{\shortstack{\textbf{Context}\\\textbf{(P50)}}} &
\multirow{2}{*}{\textbf{SA}} &
\multicolumn{3}{c}{\textbf{KV Capacity ($\rho=1/0.5$)}} &
\multicolumn{6}{c}{\textbf{Concurrency ($N_M/N_{\mathrm{SLO}}$)}} \\
\cmidrule(lr){6-8}
\cmidrule(lr){9-14}
& & & & &
\textbf{HBM (GB)} &
\textbf{SLC (TB)} &
\textbf{QLC (TB)} &
\textbf{HBM-BE} & \textbf{HBM-CDA} &
\textbf{SLC-BE} & \textbf{SLC-CDA} &
\textbf{QLC-BE} & \textbf{QLC-CDA} \\
\midrule

Qwen3-Coder & Loop
& 48 & 32.5K & 3.50
& 49.5/24.8 & 0.50/0.25 & 2.03/1.02
& \textbf{16}/240 & \textbf{5}/240
& \textbf{160}/292 & \textbf{46}/292
& 655/\textbf{292} & \textbf{187}/292 \\

& Graph
& 48 & 40.8K & 2.66
& 49.5/24.8 & 0.50/0.25 & 2.03/1.02
& \textbf{13}/189 & \textbf{5}/189
& \textbf{126}/231 & \textbf{48}/231
& 516/\textbf{231} & \textbf{194}/231 \\

\midrule

MiniMax-M2.7 & Loop
& 124 & 77.9K & 3.80
& 91.0/45.5 & 1.88/0.94 & 8.03/4.01
& \textbf{5}/249 & \textbf{1}/249
& \textbf{98}/280 & \textbf{26}/280
& 416/\textbf{280} & \textbf{109}/280 \\

& Graph
& 124 & 31.9K & 2.83
& 91.0/45.5 & 1.88/0.94 & 8.03/4.01
& \textbf{12}/593 & \textbf{4}/593
& \textbf{239}/668 & \textbf{84}/668
& 1018/\textbf{668} & \textbf{358}/668 \\

\midrule

GLM-5.1 & Loop
& 53.6 & 87.5K & 4.20
& 29.2/14.6 & 0.69/0.34 & 2.93/1.47
& \textbf{3}/656 & \textbf{1}/656
& \textbf{73}/552 & \textbf{17}/552
& \textbf{313}/552 & \textbf{75}/552 \\

& Graph
& 53.6 & 38.1K & 2.77
& 29.2/14.6 & 0.69/0.34 & 2.93/1.47
& \textbf{7}/1048 & \textbf{3}/1048
& \textbf{169}/776 & \textbf{61}/776
& \textbf{722}/776 & \textbf{260}/776 \\

\bottomrule
\end{tabular}%
}
\vspace{-2mm}
\end{table*}
}

\newcommand{\tabSix}{
\begin{table}[t]
\centering
\footnotesize
\setlength{\tabcolsep}{1.8pt}
\renewcommand{\arraystretch}{1.08}
\caption{Hardware overhead of LaFlash, including per-GPU (A100) controller metadata and per-CMOS-die circuitry.}
\label{tab:tabFive}
\begin{tabular*}{\linewidth}{@{\extracolsep{\fill}}lllc}
\toprule
\textbf{Location} &
\textbf{Component} &
\textbf{Detail} &
\textbf{Metadata (KB)} \\
\midrule
\multirow{2}{*}{Host Controller}
& CDA State Table & State, capacity & 28.3 \\
& Session Group Table & Group, lifetime & 7.5 \\
Base-die Controller & Page Mapping Table & Address, page type & 1,707 \\
\midrule
\textbf{Location} &
\textbf{Component} &
\textbf{Detail} &
\textbf{Area (mm$^2$/die)} \\
\midrule
\multirow{8}{*}{CBA CMOS Die}
& Sense Amplifiers & SA + 7 latches/BL & 87.6 \\
& BL Select & 1 HV Tr. per BL & 39.8 \\
& Row Decoders & HV block select & 9.7 \\
& Plane Buffers & 2$\times$4\,KiB SRAM/plane & 13.6 \\
& MegaPlane Control & 32-plane sharing & 2.4 \\
& Die Control & Queue, lanes & 0.1 \\
& TSV & Following~\cite{flashaccel} & 12.0 \\
& \textbf{Total} & \textbf{94.1\% of 175.6\,mm$^2$} & \textbf{165.2} \\
\bottomrule
\end{tabular*}
\vspace{-2mm}
\end{table}
}

\newcommand{\tabSeven}{
\begin{table}[t]
\centering
\footnotesize
\setlength{\tabcolsep}{3pt}
\renewcommand{\arraystretch}{1.08}
\caption{Effectiveness of LaFlash on reasoning models and multi-turn chat workloads.}
\label{tab:tabSix}
\begin{tabular*}{\linewidth}{@{\extracolsep{\fill}}llcccc}
\toprule
\textbf{Model} & \textbf{Workload} &
\multicolumn{2}{c}{\textbf{Seq. Len. (P50)}} &
\multicolumn{2}{c}{\textbf{LaFlash Speedup}} \\
\cmidrule(lr){3-4}\cmidrule(lr){5-6}
& & \textbf{Context} & \textbf{Decode}
& \textbf{vs. HBM} & \textbf{vs. SLC} \\
\midrule
Qwen3-Coder & Terminal-Bench & 17,528 & 77 & 12.5$\times$ & 3.63$\times$ \\
Qwen3-Thinking & Terminal-Bench & 18,601 & 742 & 2.42$\times$ & 1.58$\times$ \\
Qwen3-Thinking & ShareGPT & 4,357 & 1,490 & 2.19$\times$ & 1.26$\times$ \\
\bottomrule
\end{tabular*}
\vspace{-2mm}
\end{table}
}

\newcommand{\tabEight}{
\begin{table}[t]
\centering
\footnotesize
\setlength{\tabcolsep}{1.8pt}
\renewcommand{\arraystretch}{1.08}
\caption{Comparison between SSD-based KV cache retention
and LaFlash on Qwen3-Coder-30B-A3B with A100 baseline.}
\label{tab:tabSeven}
\begin{tabular*}{\linewidth}{@{\extracolsep{\fill}}lcc}
\toprule
\textbf{Parameter} &
\textbf{CachedAttention~\cite{cachedattn}} &
\textbf{LaFlash} \\
\midrule
Memory Architecture & GPU + HBM + SSD & GPU + HBM/QLC-HBF \\
\midrule
Total Capacity (GB) & 2064 & 2064 \\
NAND Peak Aggr. BW (GB/s) & 14.7 & 2048 \\
E2E Throughput (Sessions/min) &
0.0461 &
3.86 \\
\bottomrule
\end{tabular*}
\vspace{-2mm}
\end{table}
}

\makeatletter
\renewcommand{\@affiliationfont}{\fontsize{9}{10}\selectfont}
\makeatother

\begin{document}

\title{LaFlash: Capacity-Driven Acceleration for 
Coding Agent Workflows via QLC-HBF}

\author{%
\fontsize{11}{12}\selectfont
{Yuseon Choi\textsuperscript{1}, Eunjong Lee\textsuperscript{1}, Sanghyeon Park\textsuperscript{1},
Seongyon Hong\textsuperscript{2}, Sangjin Kim\textsuperscript{1,\textdagger},
and Thierry Tambe\textsuperscript{3,\textdagger}}
}

\affiliation{
  \institution{%
    {\fontsize{10}{11}\selectfont
    $^{1}$GIST, Republic of Korea;
    $^{2}$KAIST, Republic of Korea;
    $^{3}$Stanford University, USA}
  }
  \city{}
  \country{}
}

\email{%
{yuseon.choi}@gist.ac.kr,
{adria1202, shpark2623}@gm.gist.ac.kr,
sy.hong@kaist.ac.kr,
sangjinkim@gist.ac.kr,
ttambe@stanford.edu
}

\renewcommand{\shortauthors}{Choi et al.}
\renewcommand{\shorttitle}{LaFlash: Capacity-Driven Acceleration for Coding Agent Workflows via QLC-HBF}

\begin{abstract}

As large language models (LLMs) evolve toward agentic AI, coding agents are emerging as memory-intensive workloads that repeatedly reuse persistent states, including contexts and KV caches. Emerging agentic workflows, including Loop/Graph Engineering, cause state amplification through frequent active/dormant transitions, increasing memory pressure. Evicted states incur costly recomputation upon reactivation, particularly under long-context, short-decode workloads. This motivates \textit{Capacity-Driven Acceleration (CDA)}, which explicitly accounts for state amplification to allocate memory capacity between active concurrency and dormant-state retention, replacing recomputation with memory access.

In this work, we propose \textit{LaFlash}, a High-Bandwidth Flash (HBF) architecture that exploits the \underline{La}rge capacity of Quad-Level Cell (QLC)-based \underline{Flash} to realize CDA. However, QLC's long read latency demands massive read parallelism under irregular state reactivation and varying context lengths, while limited write endurance constrains placement. LaFlash introduces \textit{Session Grouping} to balance active request counts and aggregate context lengths across groups, sustaining read parallelism and block utilization through request-exclusive allocation. Furthermore, \textit{Fast-Page Allocation} exploits QLC's read-latency heterogeneity to mitigate residual inter-group imbalance and sustain bandwidth utilization. On end-to-end coding agent workloads, LaFlash achieves 7.04$\times$ higher throughput and 10.96$\times$ higher energy efficiency over SLC-HBF with batch expansion, and 2.36$\times$ higher throughput over QLC-HBF with CDA. HBM buffering further reduces unnecessary NAND writes, improving QLC write endurance. These results demonstrate the benefits of jointly optimizing state retention and physical data placement for capacity-driven acceleration.
\end{abstract}


\maketitle

\begingroup
\renewcommand{\thefootnote}{\fnsymbol{footnote}}
\footnotetext[2]{Sangjin Kim and Thierry Tambe are co-corresponding authors.}
\endgroup

\section{Introduction}

Large language models (LLMs) are rapidly evolving toward agentic AI, enabling autonomous execution of complex tasks through external tools. In particular, coding agents repeatedly inspect, modify, and verify codebases while generating and reusing persistent states such as contexts and KV caches~\cite{claudecode, hermesagent}. Emerging \textit{agentic workflows}, including Loop and Graph Engineering, extend these capabilities through iterative refinement and task decomposition, respectively.

These workflows introduce \textit{state amplification}, substantially increasing memory capacity requirements. Frequent tool calls necessitate concurrent session execution to maintain hardware utilization, while graph branching generates independent contexts and KV caches~\cite{loopengineering, graphengineering}. Consequently, both active and dormant states accumulate, increasing memory pressure beyond conventional LLM serving. Frequent state evictions then incur costly context recomputation and repeated KV-cache writes upon reactivation.

Such \textit{recomputation} is particularly expensive in coding agents due to their long-context, short-decode characteristics~\cite{tracelab, agenticcodingwild}. Recovering evicted KV caches requires repeating prefill computation, including attention operations that scale quadratically with context length, whereas retrieving existing KV caches through external-memory access (EMA) incurs only linear data transfer costs. At a 100K-token context length, recomputation latency can be three orders of magnitude higher than EMA latency on modern GPU systems. Therefore, retaining reusable states by capacity expansion becomes critical to avoiding expensive recomputation in coding agent workloads.

Conventional LLM serving engines such as vLLM and SGLang primarily exploit additional memory capacity to expand batch sizes, amortizing model weight accesses across requests and improving hardware utilization~\cite{pagedattn, sglang}. However, the long contexts of coding agents increase KV-cache traffic, which cannot be shared across requests, limiting the marginal benefits of batching through weight reuse. Meanwhile, state amplification increases the number of dormant states, causing frequent evictions and costly recomputation under limited memory capacity~\cite{kvresidency}. Once sufficient capacity is allocated to active execution, retaining additional dormant states can eliminate recomputation overhead and provide greater acceleration potential than further batch expansion (BE). We refer to this approach as \textbf{Capacity-Driven Acceleration (CDA)}. CDA is particularly effective for workloads with long contexts, short per-turn decoding, and frequent state reuse.

\textit{High-Bandwidth Flash (HBF)} provides a promising source of capacity expansion by combining NAND flash density with HBM-class aggregate bandwidth~\cite{hthree}. However, existing HBF designs primarily employ Single-Level Cell (SLC) organization to limit read latency~\cite{flashaccel, hbflex, flint}, leaving the higher bit density of multi-level flash unexploited. \textit{Quad-Level Cell (QLC)} offers 4$\times$ higher capacity than SLC, but its longer read latency demands substantially higher plane-level parallelism to sustain bandwidth~\cite{qlc, qlc_cba}. Distributing data across more planes improves read parallelism but can reduce block utilization, while request-exclusive block allocation preserves write efficiency but suffers from dormant/active transitions and context-length imbalance. This trade-off is particularly critical given QLC's limited write endurance.

In this work, we propose \textbf{LaFlash}, a QLC-HBF architecture that enables CDA by sustaining high read bandwidth while preserving write efficiency. LaFlash introduces \textit{Session Grouping} to balance active request counts and aggregate context lengths across groups while preserving request-exclusive block allocation with group-wise FIFO scheduling. It further employs \textit{Fast-Page Allocation}, exploiting QLC's inherent read-latency heterogeneity to compensate for residual inter-group parallelism imbalance. Meanwhile, newly generated KV caches are buffered in HBM and selectively written back based on reuse behavior. 

The main contributions of this work are as follows:
\begin{enumerate}[label=(\arabic*), leftmargin=*]
    \item We characterize state amplification and costly recomputation in coding agent workflows, identifying \textit{Capacity-Driven Acceleration (CDA)} as an opportunity to exploit memory capacity for state retention beyond conventional batch expansion.

    \item We identify the challenges of sustaining read bandwidth utilization while preserving block-level write efficiency, and propose \textit{Session Grouping} to address them despite dynamic request activation and context-length imbalance.

    \item We exploit QLC's inherent page-level read-latency heterogeneity as an additional dimension for read parallelism optimization, and introduce \textit{Fast-Page Allocation} to compensate for residual inter-group imbalance.
\end{enumerate}

\section{Background}

\subsection{Coding Agent Workflow}

\figOne

\textbf{Execution Structure.}
As shown in Fig.~1, coding agents have evolved from standalone LLMs to increasingly structured systems incorporating harnesses, execution loops, and graph-based workflows. \textit{Loop Engineering} repeatedly refines solutions using execution and verification feedback~\cite{claudecode, hermesagent}. LLMs invoke external tools and incorporate their outputs into the context for subsequent inference~\cite{loopengineering}. \textit{Graph Engineering} organizes tasks into interconnected nodes, where gate-controlled routing determines execution paths and node outputs are passed to subsequent nodes as context~\cite{langgraph}. Each graph node maintains its own KV cache to preserve its execution context~\cite{graphengineering}. Both workflows exhibit long-context, short-decode characteristics, accumulating extensive contexts while generating relatively few tokens per turn~\cite{freeinference_paper, agenticcodingwild, tracelab}.

\textbf{State Reuse Behavior.}
In Coding Agent workflows, a \textit{state} refers to the execution context maintained across inference and tool invocations, including conversation history, intermediate results, and their corresponding KV caches. A state is \textit{active} during LLM inference and \textit{dormant} when execution is suspended or another context is activated. Dormant states can be reactivated for subsequent inference. We characterize state reuse using the \textit{reuse interval (RI)}, defined as the elapsed time in seconds between consecutive accesses to the same state.

\subsection{Flash Memory Characteristics}

\figTwo

\textbf{Multi-Level Cell Scaling.}
As described in Fig. 2, NAND flash increases storage density by encoding multiple bits into the threshold voltage, $V_{\mathrm{th}}$, of each cell. Single-, multi-, triple-, and quad-level cells (SLC, MLC, TLC, and QLC) store 1, 2, 3, and 4 bits per cell, respectively, with $2^b$ distinguishable voltage states for $b$ bits per cell. As more states are packed into the same voltage window, narrower sensing margins increase read complexity and sensitivity to cell degradation. QLC thus provides 4$\times$ the bit density of SLC using the same cell resources, at the cost of higher read latency and reduced endurance~\cite{qlc, intelisscc21_1266, qlc_cba}.

\textbf{Read Operation.}
NAND flash reads data at page granularity by sensing cell threshold voltages against reference voltages along a selected wordline. Cells storing multiple bits require multiple sensing operations to distinguish their encoded data, increasing read latency with bit density. Unlike SLC, QLC exhibits \textit{page-level read-latency heterogeneity}, as different logical pages require different numbers of sensing operations~\cite{paca, bdp, tsp}. Furthermore, because the bandwidth sustained by a flash plane is limited by its page size and read latency, longer read latency requires proportionally more planes operating in parallel to achieve the same aggregate bandwidth. So, plane-wise data striping is critical for high-bandwidth QLC operation.

\textbf{Write and Endurance.}
NAND flash is programmed at page granularity and erased at block granularity. Repeated program/erase (P/E) operations degrade cells, limiting write endurance. Increasing bit density further reduces endurance, from approximately $100\mathrm{K}$ P/E cycles for SLC to around $1\mathrm{K}$ cycles for QLC in representative devices~\cite{dell2022qlc}. Since blocks must be erased before reuse, inefficient block utilization and data migration can increase physical write traffic. This overhead is quantified by the \textit{write amplification factor (WAF)}, defined as $\mathrm{WAF}=\mathrm{physical~writes}/\mathrm{logical~writes}$~\cite{writeperformance}. Therefore, minimizing WAF is critical for QLC-based systems.

\subsection{High-Bandwidth Flash (HBF)}

\textbf{HBF Architecture.}
HBF integrates stacked NAND dies with a logic base die, combining NAND density with high bandwidth through massive plane parallelism and wide package-level connectivity~\cite{hthree}. Recent HBF specifications define 8-high and 16-high NAND stacks, capacities up to 512 GB, and bandwidth grades of approximately 0.4--3.0 TB/s through a UCIe-based xPU--HBF interface~\cite{ocp2026hbf}. Sustaining this bandwidth requires sufficient plane-level parallelism to hide NAND read latency. For page size $P$, read latency $t_R$, and target bandwidth $B$, the required plane parallelism scales as $N_{\mathrm{plane}} \geq Bt_R/P$. QLC's longer read latency demands more planes, increasing peripheral area and power overhead. Existing SLC-based HBF therefore leaves multi-bit-per-cell density scaling unexploited.

\figThree

\textbf{HBM--HBF Integration.}
GPU memory systems can integrate HBF through HBF-only, cascaded, or parallel organizations. HBF-only suffers from write amplification due to fragmented updates~\cite{hbflex}, while cascaded HBM--HBF limits independent bandwidth utilization by sharing the HBM access path~\cite{flint}. In contrast, parallel HBM--HBF enables concurrent memory access~\cite{flashaccel, beyondcapacity}, which LaFlash adopts with HBM write buffering.

\section{Motivation}

\subsection{State Amplification in Agentic Workflows}

Coding agents retain both active and dormant states for future reuse. We define \textit{state amplification} as the increase in retained states beyond conventional active execution requirements. As illustrated in Fig.~3, Loop Engineering introduces dormancy lasting 0.1--100s through tool calls (e.g. bash, subagents, and pytest) and human intervention~\cite{tracelab, freeinference_paper, pi_trace}. To sustain throughput during these stalls, continuous batching admits additional requests while retaining suspended states. In Graph Engineering, node transitions accumulate independent execution contexts that remain dormant until revisited after several turns, typically corresponding to $\sim$10s of processing.

Fig.~4 (left) quantifies state amplification using Hermes Agent-based Loop Engineering~\cite{hermesagent} and Claude+LangGraph-based Graph Engineering~\cite{langgraph} with GLM-5.1~\cite{glm5}. For Loop Engineering, amplification is estimated as the ratio of total execution time, including tool execution and human idle periods (capped at 10 minutes), to LLM generation time. For Graph Engineering, it is measured by the number of independently retained execution contexts per session. Loop and Graph Engineering exhibit 4.2$\times$ and 3.0$\times$ amplification, respectively, indicating substantial memory capacity pressure.

Fig.~4 (right) characterizes the reuse locality of dormant states. Most reusable states are revisited within a few turns, despite remaining inactive between accesses. Unlike conventional serving, where inactive requests are often treated as eviction candidates, dormancy in coding agents does not necessarily indicate low future reuse~\cite{rethinkingkvcacheeviction}. Instead, workflow continuity creates predictable reactivation opportunities, making recently suspended states valuable retention candidates. Moreover, reuse is associated with observable workflow events, such as tool completion and graph routing, providing structural information for selective state retention.

\subsection{Capacity as an Acceleration Resource}

Conventional LLM serving frameworks, such as vLLM and SGLang~\cite{pagedattn, sglang}, adjust the number of concurrent requests according to available memory capacity, improving serving throughput through \textit{batch expansion (BE)}. Applying this conventional serving approach to SLC-HBF primarily exploits its increased capacity to accommodate more active requests. However, as discussed in Section 3.1, state amplification in coding agents leads to the accumulation of dormant states alongside the KV caches required for active execution. Consequently, prioritizing capacity for active batch expansion competes with dormant-state retention, potentially causing eviction and subsequent recomputation of reusable states.
 
In contrast, the 4$\times$ higher bit density of QLC-HBF provides an opportunity to allocate additional capacity to retaining dormant states rather than  accommodating more active requests. This is particularly valuable for coding agents, which exhibit long-context, short-decode characteristics, making recomputation of accumulated contexts substantially more expensive than retrieving retained KV caches. Once sufficient capacity is secured for the active working set, additional capacity can be devoted to state retention, replacing otherwise necessary recomputation with external-memory access (EMA). Thus, QLC-HBF enables a shift beyond conventional batching-oriented capacity scaling toward acceleration through state reuse.

\subsection{Challenges of QLC-Based HBF}
\figFour

\figFive

QLC-HBF provides the large capacity required for CDA, but QLC's long read latency poses the most fundamental obstacle to translating this capacity advantage into actual performance gains. As shown in Fig. 5, QLC thus demands substantially greater plane-level parallelism than SLC to sustain the same HBF bandwidth. However, Coding Agent workloads do not always expose sufficient memory-level parallelism. coding agents exhibit varying context lengths and execution stages across requests, while active states change dynamically with tool execution and graph transitions. Consequently, only a limited subset of stored states may be accessed concurrently, resulting in poor bandwidth utilization despite sufficient HBF capacity.

Achieving sufficient read parallelism introduces a fundamental trade-off with write efficiency due to NAND's block-granularity erase operations. \textit{Read-first mapping} stripes request data across numerous planes for interleaved prefetching. However, QLC-HBF requires more planes and pages per block than SLC-HBF, increasing the data needed to fill distributed blocks and leaving them underutilized. Conversely, \textit{write-first mapping} dedicates entire blocks to individual requests, improving write efficiency but causing plane-level workload imbalance as requests exhibit different activity and access patterns. Therefore, QLC-HBF requires a data organization that co-optimizes read parallelism, and block utilization.

\figSix

\section{Capacity-Driven Acceleration}

\textbf{State Retention over Batch Expansion.}
While conventional serving frameworks prioritize batch expansion (BE), our strategy allocates additional capacity to retaining dormant states rather than accommodating more active requests. We refer to this approach as \textit{Capacity-Driven Acceleration (CDA)}. Conventional recomputation-aware KV cache offloading introduces a low-bandwidth, high-capacity backing tier~\cite{cachedattn, mooncake}, whereas CDA expands high-bandwidth main memory itself to exploit additional capacity for acceleration. This shifts the optimization focus from whether to use a backing tier~\cite{elmoe} to how expanded main-memory capacity should be allocated between batch expansion and state retention. As illustrated in Fig. 6, conventional serving may evict dormant states to accommodate newly active requests, requiring expensive recomputation upon reactivation. In contrast, CDA retains dormant states within the expanded main memory, enabling their KV caches to be retrieved without recomputation.

\textbf{Benefits of State Retention for Long-Context Workloads.}
Coding agent workloads exhibit long accumulated contexts and short per-turn decoding, as conversation histories and tool outputs extend contexts while generating relatively few new tokens per turn. As shown in Fig.~7(left), larger batches improve compute utilization by amortizing weight EMA across requests. However, as context length increases, KV-cache traffic increasingly dominates memory bandwidth, diminishing weight amortization benefits and limiting throughput gains from further batch expansion.

Meanwhile, Fig.~7(right) shows that recomputation costs increase with context length. Recovering an evicted KV cache requires repeating prefill, including attention computation with $\mathcal{O}(L^2d)$ complexity. Accounting for linear projections and feed-forward layers, the total recomputation complexity becomes $\mathcal{O}(L^2d+Ld^2)$, whereas retrieving a retained KV cache through external-memory access (EMA) requires only $\mathcal{O}(Ld_{\mathrm{KV}})$ data transfer, where $L$, $d$, and $d_{\mathrm{KV}}$ denote context length, hidden dimension, and KV dimension, respectively. Consequently, longer contexts reduce the marginal benefits of batching while increasing the relative value of state retention.

\figSeven
\figEight

\textbf{Benefits of Replacing Recomputation with EMA.}
Frequent state reactivations in Coding Agent workflows amplify the cost of recomputation. With short per-turn decoding, this overhead remains significant even when amortized across multiple decoding turns. As shown in Fig. 8(left), recomputing a state incurs 2.7k$\times$ higher hardware cost than retrieving its KV cache through EMA. Increasing retention capacity allows more reactivations to be served through EMA, extending the interval between recomputations, as illustrated in Fig. 8(middle). To quantify this benefit, we consider a baseline where SLC-HBF accommodates the full active working set of conventional serving, while the additional capacity enabled by QLC-HBF is allocated to dormant-state retention. As shown in Fig. 8(right), this 4$\times$ capacity scaling achieves up to 7.6$\times$ speedup, demonstrating the acceleration potential of CDA.

\textbf{Active-to-Retention Concurrency Allocation.}
To realize CDA, expanded memory capacity must be allocated between active concurrency and dormant-state retention. Motivated by CONCUR~\cite{concur}, which controls KV-cache pressure to mitigate thrashing, we allocate a fraction $\rho=0.5$ of the available KV-cache capacity to concurrency provisioning. Following DistServe~\cite{distserve}, we adopt a 200-ms P90 TPOT SLO for coding workloads to constrain active concurrency. The resulting active concurrency is given by
\begin{equation}
N = \min\left(
\left\lfloor
\frac{\rho C_{\mathrm{KV}}}
{SA \cdot M_{\mathrm{KV}}}
\right\rfloor,
N_{\mathrm{SLO}}
\right),
\end{equation}
where $C_{\mathrm{KV}}$ is the capacity remaining after model-weight allocation, $M_{\mathrm{KV}}$ is the per-session KV-cache footprint, $SA$ is the workload-specific state amplification factor, and $N_{\mathrm{SLO}}$ is the estimated SLO-constrained concurrency limit.

Conventional BE approaches, especially CONCUR~\cite{concur},  implicitly assume $SA=1$, allowing the remaining $(1-\rho)$ capacity to retain evicted states and avoid some recomputation. Such implicit retention could be sufficient if state reuse were concentrated in a small set of hot states, even under high state amplification. However, our characterization reveals that coding agents exhibit broad, recurring state reuse across dormant states rather than repeatedly accessing a small hot set. Consequently, a fixed capacity margin cannot adequately preserve the reusable working set. CDA instead explicitly accounts for the observed state amplification when determining active concurrency, reserving sufficient capacity for dormant-state retention and reducing recomputation upon reactivation.

\section{QLC-HBF Architecture}

\subsection{Architecture Overview}
\figNine

\textbf{System Organization.}
Fig. 9 presents the overall architecture of \textit{LaFlash}, a QLC-based High-Bandwidth Flash system designed for Capacity-Driven Acceleration. LaFlash adopts a parallel HBM--HBF organization, where HBM provides low-latency access and write buffering, while QLC-HBF stacks provide large-capacity storage for persistent Coding Agent states and model weights. Each HBF stack comprises multiple NAND core dies interconnected through TSVs and a base die equipped with a UCIe-based host interface. Within each core die, CMOS directly Bonded to Array (CBA) technology~\cite{cba} vertically integrates NAND arrays and peripheral CMOS circuits through high-density bonding connections. Following FlashAccel, LaFlash groups planes into \textit{MegaPlanes} that share command and address control~\cite{flashaccel}.

\textbf{Architectural Constraints and Design Choices.}
QLC-HBF introduces four architectural challenges:
\textit{(1) Peripheral area overhead:} Sustaining HBF bandwidth with QLC requires more planes, increasing peripheral circuit area, particularly for sense amplifiers (SAs) and SRAM page buffers. LaFlash assesses the feasibility of accommodating these circuits using CBA integration with a 45-nm CMOS process. \textit{(2) Control overhead:} Massive plane parallelism increases address and command management overhead. LaFlash adopts FlashAccel's MegaPlane organization, grouping 32 planes under shared control, and extends it with page-type-aware scheduling to accommodate QLC's heterogeneous read operations. \textit{(3) Block utilization:} QLC increases the number of pages per block from 256 to 1,024, exacerbating the trade-off between read parallelism and block utilization. LaFlash adopts request-granular block allocation, reserving dedicated blocks for each request based on a minimum context size of 10K tokens and scaling the allocation with request length. \textit{(4) Write endurance and bandwidth imbalance:} Following FlashAccel~\cite{flashaccel}, we adopt FCR and WARM~\cite{fcr_12,warm}, relaxing retention from 3 years to 3 days for a conservative 10$\times$ endurance improvement. Originally designed for multi-level NAND, these techniques are directly applicable to QLC without modifying Vth distributions or programming voltages. HBM buffers newly generated KV caches to prevent write fragmentation and avoid unnecessary writes of short-lived or low read-to-write-ratio (below 7K) data.

\subsection{QLC Read-Latency Heterogeneity}

\figTen

\textbf{Page-Level Latency and Programming Flexibility.}
Unlike SLC, QLC exhibits heterogeneous read latency across its four logical page types due to different numbers of sensing operations~\cite{paca}. As illustrated in Fig. 10(a), Gray-code mapping of 16 threshold-voltage states enables LSB, CSB, MSB, and TSB pages to be distinguished with 1, 2, 6, and 6 sensing operations~\cite{intelisscc21_1266}, respectively. We classify LSB and CSB as \textit{fast pages} and MSB and TSB as \textit{slow pages} based on their sensing latencies. To exploit this heterogeneity, LaFlash leverages two existing programming techniques: \textit{Two-Step Programming (TSP)} separates QLC programming into intermediate and final stages, allowing logical pages to be programmed at different times under predefined dependencies~\cite{bdp, tsp}. \textit{Bidirectional Programming (BDP)} extends TSP by supporting forward and reverse programming sequences under a common Gray code, providing additional flexibility in logical-page allocation~\cite{bdp}. LaFlash exploits these mechanisms for page-type-aware placement while respecting programming-order constraints.

\textbf{Read Parallelism Optimization.}
Page-level latency heterogeneity provides an additional degree of freedom for controlling the parallelism required to sustain bandwidth. As shown in Fig.~10(b), sparse accesses to slow pages may cause bus stalls and bandwidth underutilization, while fast pages reduce the required parallelism through shorter read latency. This allows sensing and data transfers to overlap even under sparse accesses.

\subsection{Parallelism-Aware Data Mapping}
\figEleven

\textbf{Limitations of Conventional HBF Mapping.}
As shown in Fig. 11(a), FlashAccel~\cite{flashaccel} maps requests to dedicated planes, preserving write locality but suffering from context-length imbalance and idle planes when requests become dormant. HBFlex~\cite{hbflex} groups requests with similar lifetimes within blocks to improve write efficiency, but still suffers from context-length imbalance and, more critically, active/dormant imbalance as grouped requests transition independently. In contrast, Flint~\cite{flint} widely stripes each request across all planes, maximizing read parallelism at the expense of block utilization. With QLC's larger blocks and massive plane count, even a request's maximum context length may be insufficient to fill its allocated blocks, increasing write amplification. Consequently, FlashAccel, HBFlex, and Flint exhibit progressively higher read parallelism but lower block-level write efficiency.

\textbf{Session Grouping.}
LaFlash addresses this trade-off by organizing sessions into groups mapped across multiple MegaPlanes, as illustrated in Fig.~11(b). Within each group, session data follow \textit{plane-first mapping} while maintaining dedicated block ownership. Grouping targets two objectives: \textit{active-count balance}, maintaining similar numbers of active requests across groups, and \textit{context-length balance}, equalizing their aggregate active context lengths. Since sessions admitted at similar times tend to accumulate comparable context lengths, LaFlash assigns incoming sessions to groups with the fewest total sessions, without explicit length prediction. Group-wise FIFO replacement further maintains active-count balance as requests complete or become dormant. Even assuming uniform page-read latency, this organization mitigates both context-length and active/dormant imbalance while preserving request-wise block utilization.

\textbf{Fast-Page Allocation.}
Session Grouping alone cannot eliminate residual imbalance from variable context growth and state transitions. Data that can be evenly distributed across session groups is allocated to slow pages, where sufficient parallelism can hide longer sensing latency. Specifically, LaFlash balances slow-page occupancy across groups by accounting for both existing allocations and incoming writes, including active and dormant states. The remaining unevenly distributed data is placed on fast pages, whose shorter read latency reduces the required parallelism. Thus, fast pages absorb residual imbalance without requiring additional activated planes. For example, as illustrated in the second stage of Fig.~11(b), fast pages require only one-third of the plane parallelism of slow pages to sustain the same bandwidth, allowing residual accesses to maintain high bandwidth utilization even with fewer active planes.
\figTwelve

\textbf{Adaptation to Dynamic State Transitions.}
As shown in the final stage of Fig.~11(b), state transitions, such as request queuing and graph-node handoffs within a session group, can disrupt the initial balance. Group-wise FIFO scheduling mitigates the resulting active-count imbalance by coordinating subsequent session admission. Meanwhile, LaFlash continuously updates group-wise slow-page occupancy and applies Fast-Page Allocation to subsequent writes, restoring page-type balance. Since fast and slow pages share physical QLC blocks, this allocation operates within dedicated blocks reserved for every 10K tokens of context to preserve block utilization.

\subsection{Runtime Data Management}

\textbf{HBM Buffering.}
LaFlash buffers newly generated KV caches and contexts in HBM before writing them to QLC-HBF. These data may not become part of reusable prefixes depending on subsequent tool outputs and workflow transitions. While states with sufficient future reuse potential are eventually written to HBF, those belonging to sessions approaching completion can remain in HBM until termination, avoiding unnecessary QLC writes. This selective retention can accommodate states with read-to-write ratios of up to approximately 7K. Buffering also decouples incremental KV generation from HBF writes, preventing writes to individual planes from interfering with model-weight reads.

\textbf{Writeback Management.}
Fig.~\ref{fig:figTwelve} illustrates LaFlash's runtime data management architecture. LaFlash employs a host-side KV controller comprising a Writeback Controller, an Address Generator, a DMA Engine, and metadata tables to manage KV cache migration from HBM to QLC-HBF. The Writeback Controller determines when to write back buffered KV caches based on reuse eligibility and write urgency. Reuse eligibility considers prefix merging, and session maturity, while write urgency accounts for HBM capacity pressure and write batching opportunities. To minimize interference with model-weight reads, LaFlash batches writebacks and coordinates their execution across planes rather than issuing frequent writes to isolated planes.
\tabOne

\textbf{Group and Page Allocation.}
The Address Generator places selected KV caches according to the Session Grouping and Fast-Page Allocation policies described earlier. As shown in the right inset of Fig.~\ref{fig:figTwelve}, it assigns incoming sessions to groups based on the total session count and tracks per-request pending write volumes and group metadata. It then calculates the total slow-page occupancy of each group, including both active and dormant states, and distributes incoming writes to minimize occupancy differences across groups. Within each group, slow pages are assigned to individual requests based on their pending write volumes, while residual data is allocated to fast pages to mitigate remaining read-parallelism imbalance. The resulting page allocations are applied consistently across the MegaPlanes assigned to each group using request-exclusive blocks. For physical placement, the Address Generator prioritizes wordlines that have completed the first programming stage of forward TSP (FTSP) or reverse TSP (RTSP), allowing their contents to be finalized through the second stage. Finally, it generates group IDs, page types, page counts, and physical addresses for DMA transfers, while updating the Request Table, Group Table, and Page Bitmap to support subsequent state retrieval and replacement.

\section{Implementation}
\tabTwo

\textbf{LaFlash Framework.}
We developed two trace-driven simulators to evaluate LaFlash. A Python-based logical simulator tracks KV cache allocation, eviction, and reuse to isolate CDA benefits from hardware constraints. For hardware evaluation, we extended HBFSim~\cite{hbfsim_simulator} by replacing its ServeLoop frontend~\cite{serveloop} with a coding agent trace driver modeling workflow transitions, tool/human delays, and memory I/O. We customized its C++ host FTL for LaFlash's control mechanisms and incorporated QLC page-dependent read timing. Memory completion times are fed back to the driver to capture their impact on execution. Multi-GPU computation and communication overhead is modeled using NVIDIA AIConfigurator~\cite{aiconfigurator}, accounting for Tensor/Expert/Data Parallelism (TP, EP, and DP).

\textbf{System Configuration.} 
Table 1 summarizes three GPU--HBF platforms based on NVIDIA A100~\cite{nvidia_a100}, H100~\cite{nvidia_h100}, and H200~\cite{nvidia_h200}, paired with Qwen3-Coder-30B-A3B~\cite{qwen3}, MiniMax-M2.7~\cite{minimax}, and GLM-5.1~\cite{glm5}, respectively. We retain one HBM stack for A100/H100 and two for H200, replacing four stacks with HBF. Each HBF stack matches the corresponding HBM peak bandwidth while providing 8$\times$/32$\times$ SLC/QLC capacity, consistent with OCP HBF Speed Grades 1 and 2. Qwen3-Coder uses one GPU, MiniMax-M2.7 uses four-GPU TP/EP, and GLM-5.1 uses eight-GPU DP/EP.

\textbf{NAND Parameters.} 
Table 2 compares SLC- and QLC-HBF organizations for the H100 configuration, both with 8 dies per stack and 4-KiB pages, while QLC provides 4$\times$ more pages per block. We model read latency as $t_R(n)=t_{\mathrm{fixed}}+n\,t_{\mathrm{sense}}$, where $n$ denotes sensing operations. Fitting reported QLC latencies (48/76/134/228~$\mu$s at 16 KiB) under 1-2-4-8 sensing yields $t_{\mathrm{fixed}}:t_{\mathrm{sense}}\approx1:1$~\cite{date24}. An HBF patent shows that partitioning the array into shorter wordlines reduces sensing load, decreasing read latency from 15 to 4~$\mu$s as the physical page size shrinks from 8 to 4 KiB~\cite{hbnvm}. Applying the fitted ratio to this 4-$\mu$s SLC reference and adopting demonstrated 1-2-6-6 Gray coding~\cite{intelisscc21_1266} yields projected QLC latencies of 4/6/14/14~$\mu$s. Accordingly, we increase plane parallelism from 768 to 2,688 planes per stack to maintain comparable peak bandwidth.
\tabThree

\textbf{Coding Agent Workflows.} 
Table 3 summarizes Loop and Graph Engineering workloads. For Loop Engineering, we use real-world FreeInference and Pi traces~\cite{freeinference_paper,freeinference_trace,pi_trace} capturing human--agent interactions. For Graph Engineering, we construct autonomous LangGraph~\cite{langgraph} workflows using Claude~\cite{claudecode} and Hermes~\cite{hermesagent}, evaluated on Terminal-Bench 2.1~\cite{terminalbench} and SWE-bench Pro~\cite{swebenchpro}. Both workloads use Qwen3-Coder-30B-A3B, MiniMax-M2.7, and GLM-5.1 with vLLM-based serving~\cite{pagedattn}. Since Qwen3-Coder is absent from FreeInference, we augment its SWE-bench Verified~\cite{swebenchver} traces with human intervention patterns extrapolated from FreeInference. Both workloads exhibit long session contexts (106K tokens on average), short per-turn decoding (386 tokens), and 1.15 tool calls per turn on average.

\textbf{Dormancy and State Amplification.}
Fig.~13 characterizes state dormancy and reuse in coding agent workloads. Both Loop and Graph Engineering exhibit dormancy durations comparable to LLM generation time, resulting in substantial state amplification. However, their dormancy distributions differ significantly. Loop Engineering exhibits substantial latency variation due to tool execution, subagents, and human intervention, whereas Graph Engineering eliminates human intervention but introduces dormancy through multi-agent execution and node transitions, with relatively concentrated latency distributions. Dormant states with reuse intervals exceeding 10 minutes are evicted rather than retained. 

\textbf{Recomputation vs. EMA Cost.}
Table 4 analytically compares recomputation and EMA latencies across context lengths of 50K--200K tokens for the three GPU--LLM configurations. EMA latency is estimated from KV cache size and HBF bandwidth, whereas recomputation requires repeating GPU operations, including attention and linear projections, that would otherwise be avoided through KV cache reuse. We estimate these costs using GPU compute throughput and HBM bandwidth, accounting for model-specific architectures such as GQA and MLA with DSA. In multi-GPU configurations, recomputation becomes even more expensive due to additional inter-GPU communication under TP and EP. At the average context lengths in Table 3 (48K--96K tokens), recomputation incurs approximately 1,416--5,772$\times$ higher latency than EMA, highlighting the benefits of dormant KV cache retention. For A100, FP8 workloads are modeled using FP16 compute throughput, as A100 lacks native FP8 Tensor Core support.
\figThirteen
\tabFour

\tabConcurrency
\section{Evaluation}

All baselines employ vLLM-style continuous batching with first-come, first-served (FCFS) admission. Active concurrency ($N$) follows the capacity allocation policy in Section 4, using the P50 per-session KV-cache footprint with vLLM's page-granular allocation~\cite{pagedattn}. We estimate the SLO-constrained concurrency limit using an analytical performance model. We consider two capacity allocation policies, BE and CDA, which differ in whether concurrency comprises only active requests or both active and dormant requests. For physical mapping, naïve HBF baselines adopt FlashAccel's request-exclusive block mapping~\cite{flashaccel}, with naïve QLC using 3-4-4-4 sensing. We further evaluate HBFlex~\cite{hbflex}, Flint~\cite{flint}, and LaFlash under identical QLC-HBF configurations and CDA policies to isolate physical mapping benefits.

\subsection{End-to-End Benefits of LaFlash}

\textbf{End-to-End Performance and Energy Efficiency.}
Fig. 14 compares end-to-end throughput and energy efficiency, measured in sessions per second and normalized sessions per joule, respectively, across three GPU--LLM configurations. Each HBF configuration denotes the corresponding HBM--HBF hybrid system specified in Table 1. Compared with SLC-HBF+BE, which combines FlashAccel-style mapping with CONCUR-style concurrency management, LaFlash achieves an average of 7.04$\times$ higher throughput and 10.96$\times$ higher energy efficiency across the evaluated workloads.

Notably, even SLC-HBF+CDA improves throughput by 1.82$\times$ over SLC-HBF+BE, demonstrating the benefits of explicitly accounting for state amplification without increasing memory capacity. QLC-HBF+CDA further improves throughput to 3.04$\times$ by leveraging additional retention capacity. However, MiniMax-M2.7 exhibits relatively limited gains due to its larger per-token KV-cache footprint, which restricts active concurrency despite the expanded capacity. Moreover, these logical benefits are not fully realized by conventional HBF architectures, whose physical mapping schemes cannot effectively accommodate the dynamic access patterns of coding-agent workloads. LaFlash bridges this gap through session grouping and fast page allocation, achieving an additional 2.36$\times$ throughput and 1.59$\times$ energy efficiency over QLC-HBF+CDA. 
\figFourteen

\subsection{CDA Evaluation}

\figFifteen

Fig. 15 isolates the logical benefits of CDA by excluding physical mapping constraints. Even with CONCUR memory pressure management ($\rho=0.5$)~\cite{concur}, BE can retain frequently reused dormant states within its remaining capacity, partially avoiding recomputation. Such implicit retention is effective when reuse is concentrated in a small hot working set. However, coding-agent workloads exhibit broad and recurring state reuse across dormant states, making this fixed capacity margin insufficient to accommodate their amplified working sets.

As shown in Fig.~15(a), QLC-HBF+BE benefits from increased active concurrency but does not consistently translate its additional capacity into proportional throughput gains. Fig. 15(b) further reveals substantial recomputation overhead under BE, as it fails to account for state amplification despite provisioning concurrency based on estimated SLO constraints. In contrast, CDA explicitly accounts for state amplification when allocating active concurrency, reducing recomputation through dormant-state retention. Consequently, SLC-HBF+CDA achieves up to 8.62$\times$ higher ideal throughput than SLC-HBF+BE, while QLC-HBF+CDA further exploits expanded retention capacity, although its benefits vary with workload characteristics and SLO constraints. These results demonstrate the importance of balancing active concurrency and dormant-state retention rather than simply expanding the active set. Section~7.3 details LaFlash's realization of CDA's theoretical gain, and Section~8 explores optimal concurrency allocation between BE and CDA.

\subsection{LaFlash Evaluation}

\figSixteen

Fig. 16 compares FlashAccel~\cite{flashaccel}, HBFlex~\cite{hbflex}, Flint~\cite{flint}, and LaFlash under identical QLC-HBF and CDA configurations to evaluate physical data mapping. Normalized to Ideal CDA, LaFlash achieves approximately 68\% of ideal end-to-end throughput on average, compared with 38\%, 54\%, and 62\% for FlashAccel, HBFlex, and Flint, respectively. Existing mapping schemes exhibit limited bandwidth utilization or inefficient block utilization, restricting their effective throughput. In contrast, LaFlash achieves 63\% bandwidth utilization and 74\% block utilization, compared with 32\%/83\%, 46\%/79\%, and 63\%/55\% for FlashAccel, HBFlex, and Flint, respectively. LaFlash combines session grouping with fast page allocation to achieve a Pareto-optimal balance between read parallelism and block utilization, thereby achieving the highest average end-to-end throughput among the evaluated physical mapping schemes.

\subsection{Ablation Study}

\textbf{Performance and Energy Efficiency.}
Fig. 17(a) evaluates the cumulative contributions of LaFlash's optimizations. Naïve QLC-HBF reduces throughput to 0.33$\times$ that of SLC-HBF, as increased concurrency amplifies recomputation overhead. CDA mitigates recomputation by explicitly accounting for state amplification, improving throughput to 2.84$\times$, although its benefits remain constrained by FlashAccel-style mapping. Session grouping improves throughput to 5.31$\times$ by mitigating workload imbalance arising from active/dormant transitions and varying processing lengths. Fast page allocation addresses residual imbalance across session groups, ultimately achieving 6.18$\times$ throughput. Together, these optimizations translate CDA's capacity benefits into effective NAND parallelism, while improving energy efficiency by 9.98$\times$ over SLC-HBF.

Fig. 18 demonstrates the complementary benefits of session grouping and fast page allocation under realistic vLLM continuous batching. During the initial phase, frequent request arrivals introduce substantial workload imbalance, creating greater opportunities for fast page allocation to improve throughput. As serving progresses toward saturation, the relative contributions of the two mechanisms vary inversely, with fast page allocation compensating when session grouping becomes less effective, and vice versa. In particular, fast page allocation absorbs residual imbalance caused by variations in context length across session groups, sustaining throughput improvements throughout execution.
\figSeventeen
\figEighteen
\textbf{Write Endurance.}
Fig. 17(b) evaluates LaFlash's write lifetime. All evaluated configurations incorporate FlashAccel's retention-enhancement scheme (FCR/Warm~\cite{fcr_12, warm}) to improve NAND endurance. Starting from FlashAccel's SLC-HBF configuration, incorporating KV cache recomputation drastically reduces lifetime from 19.4 years to 0.29 years, as it introduces substantial NAND writes. Replacing SLC-HBF with naïve QLC-HBF slightly increases lifetime. CDA substantially improves lifetime to 8 years by reducing unnecessary KV cache re-write. Finally, HBM buffering further extends lifetime to 10.9 years (36.3\% improvement over CDA) by delaying NAND writes for short-lived data with low read-to-write ratios (e.g., below 7K for Qwen3-Coder on A100). These results highlight the need for endurance-enhancement techniques and HBM buffering to selectively retain high read-to-write-ratio states in QLC-HBF. Moreover, recomputation is fundamentally incompatible with NAND-based retention due to its substantial cumulative write amplification over long execution horizons.

\subsection{Hardware Overhead}

\textbf{Control Overhead.}
As shown in Table 6 (top), LaFlash implements its major control mechanisms through host-side memory management without modifying the GPU datapath. Session grouping, dormant-state tracking, and capacity allocation require only 11.2 KB of metadata per GPU. Meanwhile, the 1.5 MB Page Mapping Table is maintained by HBF/HBM base-die controllers, enabling local address translation and page-type-aware scheduling without centralized mapping storage.

\textbf{NAND Peripheral Area Overhead.} 
As shown in Table~6 (bottom), LaFlash preserves the NAND cell array organization and modifies only the peripheral circuitry on the CBA CMOS die. We verify the control logic in RTL against a behavioral QLC array model reproducing 4/6/14/14\,$\mu$s page-dependent latencies, and synthesize it using 45-nm technology~\cite{nangate45,freepdk45} at 400\,MHz. The page buffer, including sense amplifiers, data latches, and bitline selects, dominates the area at 77\%, whereas the control logic accounts for 1.5\%. The page-buffer area scales linearly with concurrently sensed bitlines, making QLC require at least 3.5$\times$ more area than SLC to sustain equivalent bandwidth. Plane-buffer SRAM is estimated using CACTI~\cite{cacti}, with less than 3\% error against a measured 45-nm SRAM macro~\cite{45nmsram}. Overall, the CMOS die occupies 165\,mm$^2$, fitting within 94\% of the 175.6\,mm$^2$ OCP HBF device footprint~\cite{ocp2026hbf}.

Prior HBF work estimated peripheral area by scaling NAND-process circuitry from die photographs~\cite{flashaccel}, which does not capture the potential benefits of implementing peripheral circuits in a separate CMOS process, as enabled by CBA. High Voltage(HV) transistors in CMOS Die must sustain the NAND program voltage (about 20--25\,V~\cite{toshiba-patent}), which is set by the cell and unaffected by CBA bonding. We size them from the 3.3-V device rules of~\cite{icsprout55}, with a 1.2-$\mu$m channel~\cite{macronix-patent} and voltage-scaled drain offset and isolation, yielding 3.6\,$\mu$m$^2$ bitline-select and 8.9\,$\mu$m$^2$ wordline-pass transistors (10$\times$ and 25$\times$ a minimum 3.3-V thick-oxide transistor). However, configurations with longer read latency, such as 1-2-4-8 Gray coding, require more planes and may exceed the available footprint, highlighting the need for careful co-design of page encoding, read latency, and peripheral area.
\tabSix
\tabSeven

\section{Discussion}

\textbf{LaFlash Applicability.}
LaFlash is particularly suited to coding agents, where frequent state transitions and high recomputation costs favor main-memory capacity expansion. Table 7 further evaluates workload characteristics and throughput benefits across Qwen3-Coder-30B-A3B and Qwen3-30B-A3B-Thinking-2507~\cite{qwen3}. Compared with Qwen3-Coder, Qwen3-Thinking exhibits 9.6$\times$ longer decoding sequences on Terminal-Bench, increasing weight amortization and reducing the relative benefits of dormant-state retention. Consequently, LaFlash achieves 2.42$\times$ and 1.58$\times$ higher throughput than HBM and SLC-HBF, respectively, compared with 12.5$\times$ and 3.63$\times$ for Qwen3-Coder. Conventional multi-turn workloads~\cite{vicuna2024} further limit CDA's benefits through lower state amplification, shorter contexts, and longer decoding phases. Even under these conditions, LaFlash achieves 2.19$\times$ and 1.26$\times$ higher throughput than HBM and SLC-HBF, respectively, on ShareGPT.

\textbf{Effectiveness Compared with SSD Retrieval.}
Both SSD-based KV-cache offloading~\cite{cachedattn, mooncake} and LaFlash expand KV capacity beyond HBM. However, SSDs rely on bandwidth-limited GPU--SSD interfaces, whereas LaFlash integrates QLC-HBF as a high-bandwidth memory tier alongside HBM. At comparable capacities, LaFlash achieves 84$\times$ higher end-to-end throughput, as shown in Table~8.

\tabEight

\textbf{QLC-HBF Feasibility and Future Directions.} 
QLC-HBF remains challenging to deploy without endurance-enhancement techniques, such as FCR and WARM~\cite{fcr_12, warm}, which relax retention requirements to extend NAND lifetime, as adopted by FlashAccel~\cite{flashaccel}. Practical implementations must address manufacturing-level challenges, including peripheral area and power overhead, as well as device-level reliability issues during operation, such as read-retry and retention degradation~\cite{readretry}. Future work includes heterogeneous MLC/TLC organization, pseudo-SLC-based optimization~\cite{pSLC}, and device-level endurance enhancement~\cite{qlc_womv} to improve the practicality of high-density HBF.

\section{Conclusion} 

This paper presents LaFlash, a QLC-based High-Bandwidth Flash architecture that turns memory capacity expansion into an acceleration opportunity for coding agent workloads. By exploiting frequent active/dormant transitions and broad state reuse, LaFlash introduces Capacity-Driven Acceleration (CDA), explicitly accounting for state amplification to balance active concurrency and dormant-state retention beyond conventional batch expansion. To translate QLC's high storage density into effective system performance, LaFlash combines Session Grouping and Fast-Page Allocation to sustain read parallelism while preserving block utilization under dynamic state transitions and context-length imbalance. Furthermore, HBM buffering reduces unnecessary NAND writes, mitigating QLC's endurance limitations. Across three GPU--LLM configurations and two coding agent workflows, LaFlash achieves up to 10.26$\times$ higher end-to-end throughput and 17.21$\times$ higher energy efficiency than SLC-HBF with batch expansion, realizing approximately 68\% of ideal CDA throughput on average. These results demonstrate that effectively scaling memory for agentic AI requires jointly optimizing state retention, physical data placement, and write management, rather than simply increasing storage capacity or batch size.


\bibliographystyle{ACM-Reference-Format}
\bibliography{refs}

\end{document}